\documentclass[%
 aip,
 amsmath,amssymb,
 reprint,%
]{revtex4-2}

\usepackage{graphicx}
\usepackage{dcolumn}
\usepackage{bm}
\usepackage[dvipsnames]{xcolor}
\usepackage[utf8]{inputenc}
\usepackage[T1]{fontenc}
\usepackage{mathptmx}
\usepackage{etoolbox}
\usepackage{textcomp}
\makeatletter
\def\@email#1#2{%
 \endgroup
 \patchcmd{\titleblock@produce}
  {\frontmatter@RRAPformat}
  {\frontmatter@RRAPformat{\produce@RRAP{*#1\href{mailto:#2}{#2}}}\frontmatter@RRAPformat}
  {}{}
}%
\makeatother
\begin{document}

\preprint{AIP/123-QED}
\title{Instability and stress fluctuations of a probe driven through a worm-like micellar fluid }
\author{Abhishek Ghadai}
\affiliation{Raman Research Institute, Bengaluru 560080, India}
\author{Pradip Kumar Bera}
\affiliation{Raman Research Institute, Bengaluru 560080, India}
\affiliation{University of Wisconsin-Madison, Madison, Wisconsin 53706, United States}
\author{Sayantan Majumdar*}
\affiliation{Raman Research Institute, Bengaluru 560080, India}
\email{smajumdar@rri.res.in}
\date{\today}

\begin{abstract}
A particle moving through a worm-like micellar fluid (WLM) shows instability and large fluctuations beyond a threshold \textcolor{black}{velocity/force (depending on the control parameter used)}. Despite many detailed studies, a direct measurement of the time-dependent stress on the probe particle remains unexplored. To address this, we have designed a measuring geometry coupled with a commercial rheometer to study the dynamics of a cylindrical probe through a WLM system of 2 wt.\%  cetyltrimethyl ammonium tosylate(CTAT) + 100 mM sodium chloride(NaCl) for a wide range of velocity and stress scales. We map out the in-situ velocity distribution using particle imaging velocimetry. Beyond a velocity threshold, we observe large stress fluctuation with gradual stress build-up followed by sudden stress drop indicating storage and release of elastic energy. The length scale constructed from the stress build-up time scale and the probe's velocity match the length scale of extensile deformation just before the stress drop, confirming the strong correlation of storage and release of energy with the unstable probe motion. Interestingly, the  Weissenberg number ($Wi$) for the onset of flow instability obtained from the shear and extensile components remains almost the same. We also find that the turbulent motion of the probe at higher $Wi$ results from the complex mixing of the stick-slip events originating from the partial release of the stored elastic energy. Further, we show that the magnitude of the stick-slip events depends on the detailed micellar structure and dynamics controlled by salt concentration and temperature.

\end{abstract}

\maketitle

\section{Introduction}

The motion of a particle moving under a constant applied force through a viscous fluid attains a terminal velocity when the viscous drag completely balances the applied force. Such dynamics leading to terminal velocity are well understood\cite{stokes1851effect}. On the other hand, particle motion through visco-elastic worm-like micellar (WLM) fluids can be quite complex, and terminal velocity may not exist due to various fluid instabilities\cite{jayaraman2003oscillations,chen2004flow,kumar2012oscillatory,wu2021linear}. Understanding such complex motion in WLM systems is important not only for a wide range of applications in oil drilling, hydraulic fracturing, and manufacturing cosmetic products, but they also present interesting challenges from the fundamental physics perspectives \cite{bewersdorff1988behaviour, ezrahi2007daily, dreiss2017wormlike, li2020rheological}. 

Worm-like micelles (WLMs) are long polymer-like structures formed by the self-assembly of surfactant molecules in aqueous solution. In the semi-dilute regime, the density of these polymer-like structures becomes high enough to form entanglement between them. Such entangled structures give rise to long-range correlations in the system resulting in visco-elasticity. Despite many similarities between these systems and polymeric liquids, the major difference between these two classes of systems stems from the fact that WLM can break and recombine due to thermal fluctuations giving rise to additional relaxation modes. This results in Maxwellian behaviour in a wide range of WLM systems with a well-defined relaxation time scale \cite{cates1987reptation}. Non-linear flows of WLM systems have been extensively studied using shear rheometry as they exhibit striking flow instabilities, such as shear banding, elastic turbulence, rheochaos at substantially low values of Reynolds numbers $Re$ but at high Weissenberg numbers $Wi$ \cite{salmon2003velocity,helgeson2009relating,mohammadigoushki2016flow,fardin2010elastic,majumdar2011universality,ghadai2023origin,bandyopadhyay2000observation,ganapathy2006intermittency,ganapathy2008spatiotemporal}. Such instabilities are driven by elasticity rather than inertia. Nonetheless, interesting inertio-elastic flow instabilities have also been reported in these systems when both $Wi$ and $Re$ becomes comparable \cite{perge2014inertio,mohammadigoushki2017inertio}. Although these studies provide valuable insights regarding the coupling of microstructure and flow in WLM systems, such rheological studies are inadequate to predict/describe the complex motion of a probe through WLM fluids mostly due to the wide differences in flow geometries.

Earlier experimental studies have reported the unstable motion of a sphere settling under gravity through WLM fluids. These include the study by Jayraman and Belmonte\cite{jayaraman2003oscillations} using CTAB-NaSal based WLM fluid. They observe that beyond a certain value of Deborah number $De \approx$ 45, the velocity of the sphere shows large fluctuations. They have attributed the origin of such fluctuations to the formation and breakage of shear-induced structures around the sphere. More recent studies by Mohammadigoushki and Muller\cite{mohammadigoushki2016sedimentation} and Zhang and Muller\cite{zhang2018unsteady} point out the role of extensional deformation in the wake of the moving sphere giving rise to instability beyond the certain value of the extensional Weissenberg number. Using in-situ flow birefringence studies Chen and Rothstein\cite{chen2004flow} suggest that the flow instability originates from the flow-induced rupture of stretched micelles at the rear end of the moving sphere. They also directly observe filament rupture using extensional rheometry.  
The experimental investigation of Wu and Mohammadigoushki\cite{wu2021linear} further supports a similar mechanism of micellar breakage. Kumar et al.\cite{kumar2012oscillatory} show unstable flows in CTAT + NaCl based WLM system where they report a periodic burst of damped oscillations in the sphere's velocity. The nature of such oscillations is quite distinct from the similar studies mentioned above. This implies that the microscopic micellar structure plays an important role in determining the nature of instability. Nonetheless, the steep rise in velocity of the settling sphere at the onset of a burst also suggests a sudden structural breakage. Interestingly, a recent bulk rheological study on the CTAT-NaCl WLM system highlights the role of micellar breakage in the elasticity-driven Taylor-Couette instability in these systems\cite{ghadai2023origin}. Recent numerical simulation based on two-species VCM model\cite{vasquez2007network} for WLM also supports the micellar rupture in the wake of a moving sphere as the mechanism of instability\cite{sasmal2021unsteady}. The role of surface roughness of the settling sphere has also been investigated in WLM systems\cite{sasmal2022effect}. The structure of CTAT-based WLM systems is complex and highly sensitive to the nature and concentration of added salt which in turn affect their rheological response significantly\cite{calabrese2015rheology, rassolov2022kinetics}. Very recent studies indicate that WLM systems having almost identical linear rheological properties can display significantly different non-linear dynamical behaviour\cite{pasquino2023startup}.

Despite these detailed studies, a direct measurement of temporal stress on the probe during the transition from stable to unstable motion by systematically varying the flow rate remains unexplored. Based on pressure drop measurements in an array of cylinders \cite{moss2010flow} and position fluctuations of a flexible elastomer \cite{dey2018viscoelastic} placed in a flow-field of WLM fluids, the role of stress variation during the flow instabilities have been indicated. Nonetheless, stress measurement with high spatial and temporal resolution is crucial to verify the proposed mechanisms of micellar scission in giving rise to unstable motion through WLM systems. Furthermore, such controlled experiments are required to understand the connection of the instability near the onset point to the chaotic or turbulent motion at much higher values of $Wi$ as reported in the earlier studies for CTAT-NaCl systems.

In this work, we use a home-built setup coupled with a commercial rheometer to study the stress response of a cylindrical probe driven through the WLM system with constant velocity. Our setup not only mimics the motion of small probes settling under gravity in a more controlled manner but also has other significant advantages: For a given WLM system, the wide range of velocity/force scales spanning several decades that can be probed/measured in our pin-driving setup can not be matched in conventional ball-drop experiments. Similarly, in our setup the stress fluctuation data can be acquired, as long as the sample remains stable. This allows us to perform accurate statistical analysis over very long  time scales. Thus, our setup can potentially verify the applicability of the existing macroscopic rheological models of flow instability, transient dynamics, shear banding etc.\cite{mohammadigoushki2019transient, perge2014surfactant, divoux2016shear} in the context of unstable motion of a small probe through similar systems.
This setup also allows us to probe the in-situ local deformation field using PIV and light scattering measurements. We observe the formation of a turbid band in the wake of the probe, just below the onset velocity of the probe corresponding to large stress fluctuations. We find that such a turbid band corresponds to an extended wake and its attachment-detachment dynamics with the probe is strongly correlated with the nature of the stress fluctuations.

\section{Materials and Methods}

The samples are prepared by adding known amount of cetyltrimethyl ammonium tosylate (CTAT) and sodium chloride (NaCl) in filtered deionized water followed by a through mixing inside an air-tight container. After this, the samples are kept in an incubator at $40\,^{\circ}C$ (well above the Kraft temperature) for at least 12 hours to ensure complete dissolution of the surfactant in the solution. Further, the solution is equilibrated at room temperature for a week before performing the experiments. As the micellar solution is transparent, to perform particle imaging velocimetry (PIV) measurements, we disperse a known amount (1.5 wt\%) of polystyrene tracer particles synthesized in our lab\cite{dhar2019signature} having a size distribution between 20-50 $\mu m$ in filtered deionized water before adding CTAT and NaCl. The micellar solution embedded with polystyrene beads is then incubated at $40\,^{\circ} C$ for at least 12 hours followed by equilibration at room temperature for a week. The sample is thoroughly mixed before use to ensure a uniform dispersion of the tracer particles. We verify that the addition of 1.5 wt\% tracer particles does not change the rheological behavior of the WLM system under study [Fig. S1].

In our experiments, we use a custom-made geometry that is coupled to a MCR-702 Twin-Drive stress-controlled rheometer (Anton-Paar, Graz, Austria ). The geometry has two parts:  the measuring geometry and the sample cell. The measuring geometry [Fig. 1(a)-(b)] consists of a millimeter-sized pin (typical length $\sim$ 15-20 mm, thickness\textcolor{black}{/diameter ($d_p$)} $\sim$ 1-5 mm) which is connected to the upper drive of the rheometer with the help of a compatible shaft. The sample cell is made from two co-axial cylinders \textcolor{black}{(inner radius ($r_i$) = 58 mm, outer radius ($r_o$) = 83 mm, depth = 18 mm)}. The pin can be driven inside the sample cell with either force-controlled or velocity-controlled modes (see movie 1) with the help of the rheometer. The samples are filled in the annular region between the cylinders up to a height of 15 mm. We have verified the variation in sample filling height (we have also checked for 10 mm) does not significantly affect the force response of the sample [Fig. S2]. The gap between the end of the cylinder and the bottom plate is approximately 10 mm. Interestingly, we observe that changing the length of the submerged portion of the pin by varying the gap between the bottom plate and the end of the cylindrical probe does not change the nature of the $F$ vs $v$ curves. However, increasing the length of the submerged portion causes an upshift of the $F-v$ plots due to the increasing area of contact [Fig. S3]. In our experiments, we have kept the gap between the curved surface of the pin from the side walls (inner and outer walls of the concentric cylinders as shown in Fig.1(a) and 1(b)) also 10 mm. Such an experimental design is compatible both in single drive (only the pin is moving) and in twin drive rheometric configurations (either the pin or the sample cell or both moving). In a single drive mode, the sample cell is mounted on a temperature-controlled Peltier plate. The uncertainty in sample temperature (measured directly using a thermometer) in such configuration is around $0.3\,^{\circ} C$. The pin is driven inside the sample kept at different temperatures ($20\,^{\circ} C-30\,^{\circ} C $) in order to get the temperature-dependent mechanical response of the viscoelastic fluid. The entire geometry is kept inside a humidity-controlled box to avoid sample evaporation. 

For in-situ imaging, we use a twin drive configuration where the sample holder is transparent and can be rotated at a desired velocity with the help of the bottom motor of the rheometer [Fig. S4 and S5]. The pin is held fixed in the lab frame for imaging the fluid around the pin easily. In our set-up direct temperature control of the sample is not possible in such a configuration. We can only control the ambient temperature using an air conditioner (set to $25\,^{\circ}C$) in the lab. This indirect control increases the uncertainty in sample temperature to around $1\,^{\circ} C$ (again confirmed with direct temperature measurement using a thermometer). However, we find that the nature of the mechanical response remains the same within experimental uncertainties. To observe the sample turbidity under imposed flow, we use a transparent sample cell to image the system in the transmission mode in a twin-drive configuration (see Fig. S4 for a schematic of the set-up). As illustrated in the schematic, the illumination is aligned slightly off-axis to the camera position to ensure the collection of only scattered light. To map the local velocity field, we image the sample-air interface (see Fig. S5 for a schematic of the set-up) seeded with tracer beads with the help of a fast camera (Phantom Miro C210). Such images are further analyzed using particle imaging velocimetry algorithm\cite{Thielicke_2021} (PIVlab) using MATLAB software to directly map the flow field. PIV analysis is performed by taking a square grid size of 64 pixels and an overlap of 20\% between the adjacent grids. The velocity magnitude is calculated from the x and y components followed by a fixed smoothening protocol. There is no further time-averaging to get the instantaneous velocity distribution using PIV.

\begin{figure}
    \centering
    \includegraphics[height=15 cm]{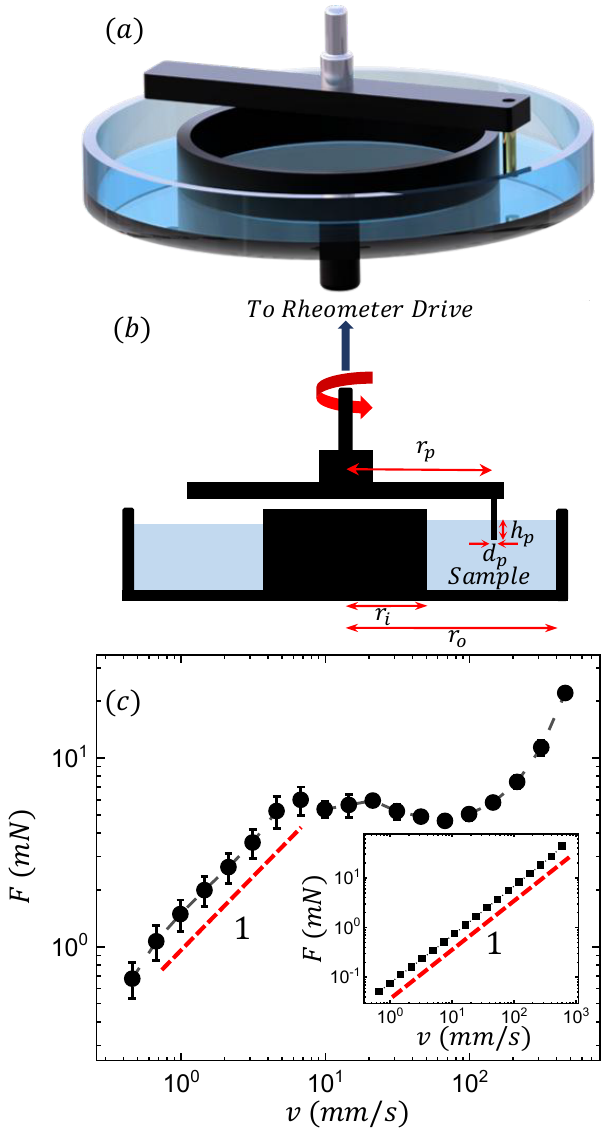}
    \caption{\label{fig:wide} (a) and (b) are the schematics of the experimental set-up of the PIN driven inside an annular sample cell. (c) $F$ vs $v$ curve at controlled velocity for a probe with 5 mm diameter at $25\,^{\circ}C$ in the case of 2 wt\% CTAT + 100 mM NaCl wormlike micellar solution. Inset shows the F-v curve for Newtonian fluid glycerol.   }
\end{figure}

The extensional behaviour is characterized by DOS (Dripping onto a substrate) measurements where a liquid bridge formed between a dispensing needle and a substrate undergoes uniaxial extensional deformation and self-thinning. For the experiments, we follow standard protocols mentioned in the literature \cite{dinic2015extensional,dinic2017pinch}. However, we have replaced the bottom glass substrate with a stainless steel plate of 4 mm diameter to pin the contact line only to the edges of the bottom substrate and avoid continuous spreading of the sample during thinning of the liquid bridge as reported by Omidvar and coworkers \cite{omidvar2019detecting}. In our experiments, the micellar solution is pumped at a flow rate of 2 $\mu l⁄min$  with the help of a syringe pump (NE-8000, New Era Pump Systems, USA) and the pumping is stopped before the droplet from the nozzle comes in contact with the substrate. We use a nozzle having an outer diameter of 1.3 mm and an aspect ratio (ratio of the gap between the nozzle and the substrate and nozzle diameter) of 3. The images are recorded at 100 frames per second with the help of a fast camera (Phantom Miro C210) and the data is analyzed using custom written codes in MATLAB software. The experiments are repeated several times to verify the reproducibility of the data.

\section{Results and discussions}

\begin{figure*}
    \centering
    \includegraphics[height=9 cm]{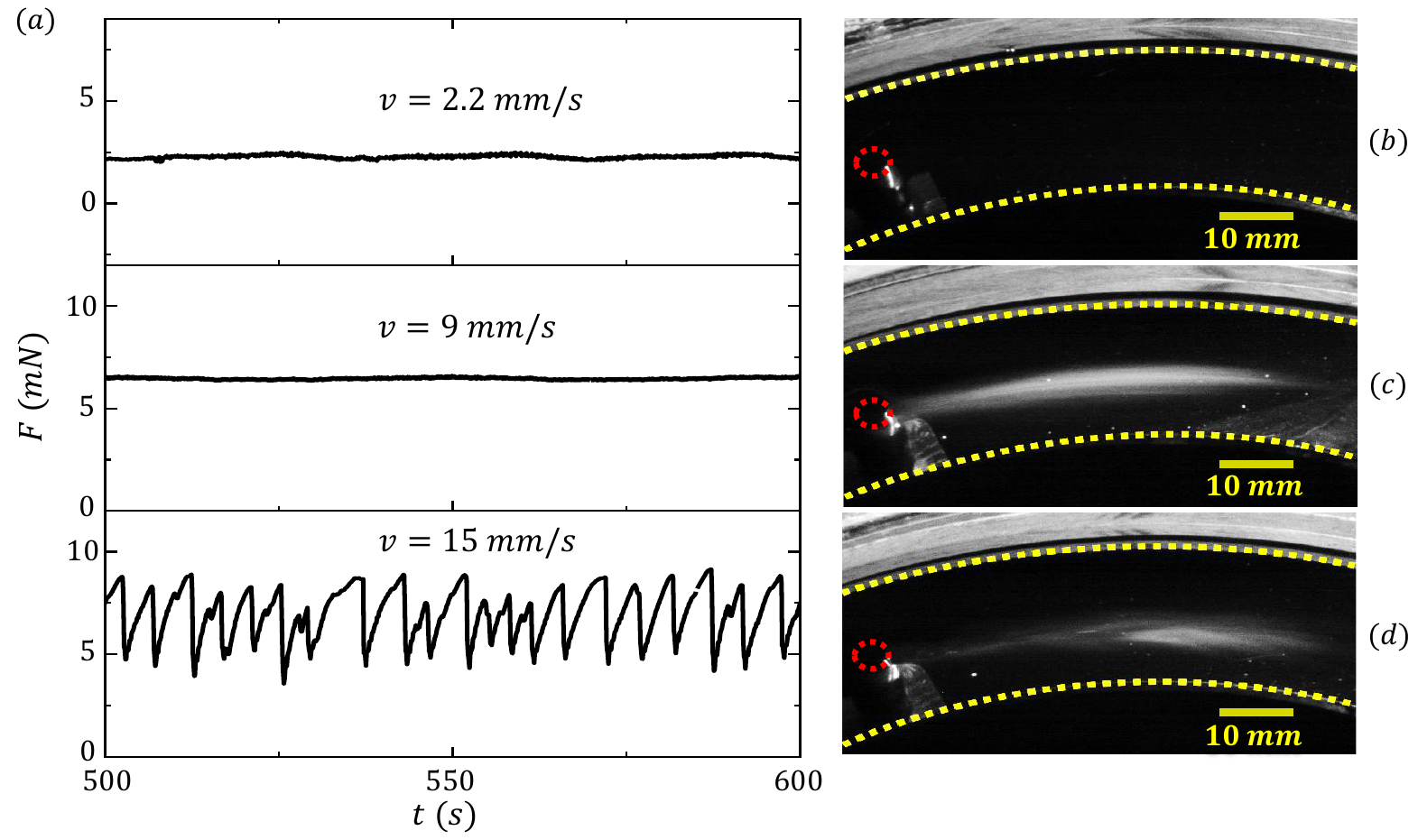}
    \caption{\label{fig:wide} (a) Temporal variation of $F$ at constant applied velocities of 2.2 mm/s, 9 mm/s, and 15 mm/s respectively at room temperature of $25\,^{\circ}C$ for 5 mm pin diameter in the case of 2wt\% CTAT + 100 mM NaCl system. Panel (b), (c), and (d) show the instantaneous snapshots of the flow-induced turbidity measurements corresponding to applied velocities of 2.2 mm/s, 9 mm/s and 15 mm/s respectively. Red dotted circle represents the pin position. }
\end{figure*}

From the raw torque ($\Gamma$) and angular velocity ($\omega$) signal we obtain the force ($F = \Gamma / R$) and velocity ($v = R \omega$). Here, $R$ corresponds to the pin's radial distance from the rotation axis as depicted in Fig. 1(b). Fig. 1(c) shows the steady state $F$ vs $v$ data for 2 wt.\% CTAT + 100 mM sodium chloride sample at $25\,^{\circ}C$ for pin diameter of 5 mm. We find that $F$ varies linearly with $v$ up to $v$ $\approx$ 10 $mm/s$. At higher velocities, we obtain a plateau region where $F$ remains nearly constant with increasing $v$. Beyond this region, we again obtain a linear variation of $F$ vs $v$ for $v >$ 100 $mm/s$. The nature of the $F(v)$ remains the same for different pins with diameters ranging from 1.5 mm to 5 mm [Fig. S6] except for an upward shift of the curves with increasing pin diameter. In Fig. 1(c) we use 10 s waiting time per data point. We further verify that for larger waiting times $> 10 s$, the $F$ vs $v$ curves remain stationary, indicating the steady state nature of $F$ vs $v$ data [Fig. S7]. \textcolor{black}{However, there is a slight deviation of $F$ at the plateau onset over repeated measurements. This could happen due to the presence of large fluctuations as discussed in the subsequent sections of the paper. }  We also measure the $F$ vs $v$ for Newtonian fluid glycerol, where we obtain a linear variation of $F(v)$ over a large range of applied velocities [inset of Fig. 1(c)] as expected. We find that $F/v$ for the micellar system is much higher than that for Glycerol for lower pin velocities [Fig. S8]. With increasing velocity beyond the plateau onset, $F/v$ shows a strong decrease due to shear-thinning. Finally, beyond the force-plateau $F/v$ micellar system becomes slightly less than that for glycerol. At such high velocities, the inverse of the shear-rate becomes smaller than the micellar breaking-recombination time scale \cite{ghadai2023origin}. This indicates that the broken micellar structures cannot re-join over such a small time scale showing a quasi-Newtonian behaviour (linear $F/v$ vs $v$ relation) above the force plateau.

\textcolor{black}{Interestingly, earlier studies on Taylor-Couette flow of CTAT-NaCl systems report the absence of a flat stress plateau \cite{ganapathy2006intermittency,majumdar2011universality,ghadai2023origin}. The significant power-law slope ($n \sim$ 0.3) at the stress plateau indicates that the CTAT-NaCl system is not expected to exhibit shear banding.} We perform bulk rheological measurements in Taylor-Couette geometry with the CTAT-NaCl WLM system by varying the salt concentrations and
temperature. We observe the absence of a flat stress plateau
in all cases (Fig. S9 and S10) over the broad parameter range probed further supporting the non-banding nature of the system. The evidence of stable shear bands is also not observed during flow-induced turbidity measurements in the gradient-vorticity plane \cite{ghadai2023origin}. However, from the PIV studies in the flow-gradient plane, we find that in the early plateau region, before
the onset of elastic turbulence, the velocity profiles across the shear gap show a time variation with the velocity fluctuations predominantly confined close to the rotating inner cylinder \cite{ghadai2023origin}. This observation suggests that the possibility of flow-induced
spatial heterogeneities in the system cannot be ruled out, although no steady shear bands are observed. On the other hand, the presence of a flat force plateau is robustly established with multiple measurements and also by varying the salt concentrations/temperature in the pin-driving setup [Fig. S9, S10 and S11].  Further, we calculated the stress due to the pin ($\sigma_{pin}$) by dividing the force over the pin contact area with the sample and shear rate ($\dot{\gamma}$) by dividing the driving velocity over the pin thickness. From the comparison of the flow curves obtained using the pin-driving setup with that observed in bulk rheology, we find a significant mismatch in the nature of the flow curves, as well as,  the stress scales [Fig. S12]. This suggests that the nature of the flow and reorganization of WLM around the cylindrical probe is quite different from that of the conventional Taylor-Couette flow and needs more detailed investigations. 
\newline
\begin{figure*}
    \centering
    \includegraphics[height=10.5 cm]{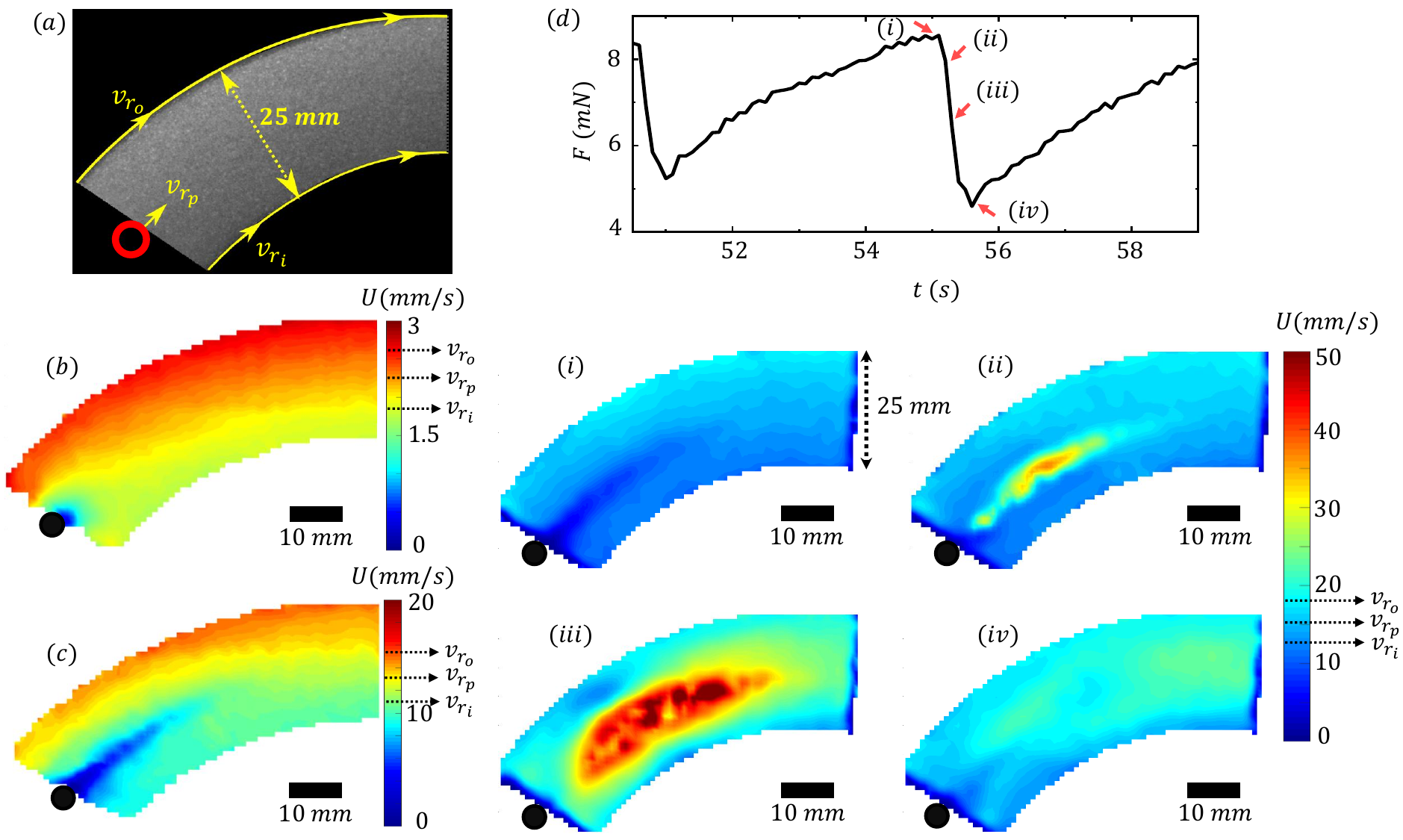}
    \caption{\label{fig:wide} (a) Raw image of the sample-air interface seeded with polystyrene tracer particles. Red circle depicts the position of the pin. $v_{r_p}$, $v_{r_i}$, $v_{r_o}$ are marked as the driving velocity at the probe position, inner wall and outer wall of the sample cell respectively.  Panel (b) and (c) show spatial velocity distribution at a particular instant of time for driving velocities of 2.2 mm/s and 12 mm/s respectively at room temperature of $25\,^{\circ}C$ in the case of pin having 5 mm diameter for 2wt\% CTAT + 100 mM NaCl system. Here, the black circle is depicted as the pin position. Zoomed in view of recorded force fluctuation [panel (d)] response in the steady-state for $v$ = 15 mm/s at room temperature of $25\,^{\circ}C$. Different regions for a force drop event are marked as i, ii, iii, and iv. Corresponding velocity field obtained using PIV are shown below. Colorbars show the magnitude of the mapped velocities in mm/s. }
\end{figure*}

\begin{figure*}
    \centering
    \includegraphics[height=12 cm]{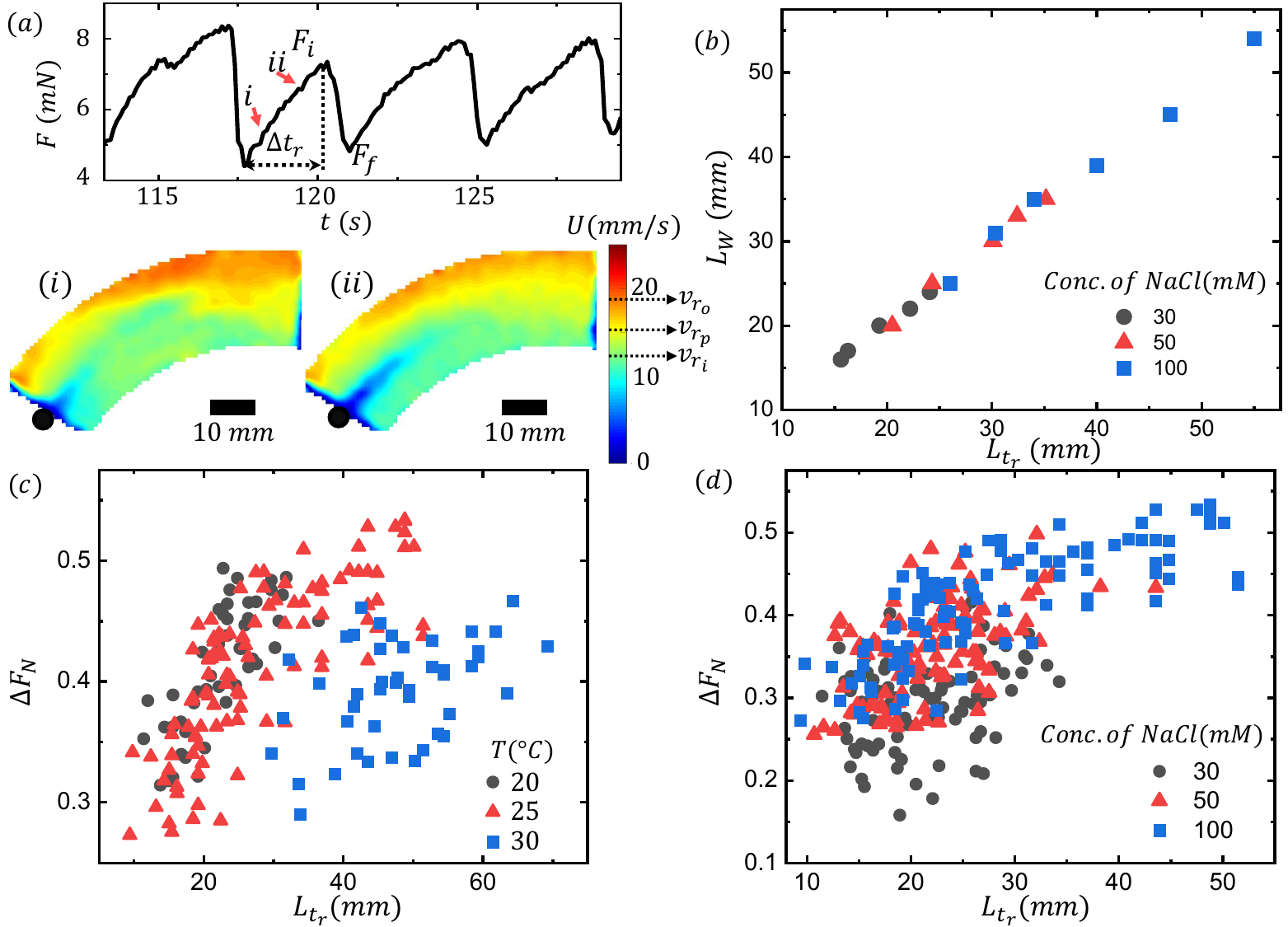}
    \caption{\label{fig:wide} (a) Two different regions during force build up are marked as i and ii in the time series and corresponding velocity distribution is shown below for a driving velocity of 15 mm/s at room temperature of $25\,^{\circ}C$ in the case of 2wt\% CTAT + 100 mM NaCl system. Colorbar represents velocity magnitude. Black circle indicates the pin position. $v_{r_p}$, $v_{r_i}$, $v_{r_o}$ are marked as the driving velocity at the probe position, inner wall and outer wall of the sample cell respectively. (b) Comparison of effective length scale of the wake ($L_{W}$) obtained from PIV measurement and time series ($L_{t_r}$) of force fluctuations for 2wt\% CTAT with varying NaCl concentrations. Normalized force drop ($\Delta F_N$) as a function of wake length ($L_{t_r}$) obtained from the time series data for varying temperature of 2wt\% CTAT + 100 mM NaCl system is shown in panel (c). Similar plot is obtained for different salt concentrations at room temperature shown in (d). }
\end{figure*}

Next, we probe the steady-state flow behaviour of the system. We show the temporal evolution of $F$ at a constant $v$ in Fig. 2(a) for three different imposed velocities: 2.2 mm/s, 9 mm/s, and 15 mm/s. For $ v = $15 mm/s which corresponds to the early plateau region of $F-v$ plot, $F$ shows large steady-state fluctuations with saw-tooth features, whereas, no significant fluctuations are observed for other values of $v$ corresponding to the linear region of $F$ vs $v$ plot. Interestingly, the saw-tooth nature of fluctuations is observed only at the early plateau region. The nature of the fluctuations becomes more complicated and chaotic deep inside the plateau region [Fig. S13]. Also, the temporal fluctuation of force reaches a steady state fairly quickly across all the driving velocities probed here and does not show any long transients [Fig. S14]. \textcolor{black}{To rule out long transients in the time series and viscous heating, we divide the force time series (obtained for at least 2000 seconds for each driving velocity) in to 4 equal parts (500 seconds in each part) and plot the mean force with error bars indicating the standard deviations [Fig. S15]. We find that the mean and the standard deviation remain the same across the different partitions of the time series. This indicates statistical stationarity, implying a steady state behaviour.} We further quantify the strength of fluctuations by calculating the standard deviation normalized by the mean and find that the strength of fluctuations is maximum exactly at the onset of the plateau region and decreases gradually towards the deeper plateau regime [Fig. S16]. We also confirm that such fluctuations are not observed for Newtonian fluid glycerol over the whole range of $v$ that we probe for CTAT-NaCl system [Fig. S17]. This highlights the important role of underlying micellar microstructures and dynamics in generating such unstable flows. 
To gain insights into the origin of the unstable flows in the system, we study the flow-induced sample turbidity using in-situ transmission-mode imaging [Fig. S4]. In WLM systems, sample turbidity occurs due to shear-induced local density variation over a range of length scales, enabling visible light scattering\cite{chen2004flow, fardin2009taylor,ganapathy2008spatiotemporal2}. We perform these experiments in the separate motor transducer (SMT) mode where the pin is held fixed and the bottom sample cell is rotated at a desired angular velocity with the help of the bottom motor. Figs. 2(b)-(d) show the snapshots of the sample turbidity in steady-state corresponding to fixed applied velocities of 2.2 mm/s, 9 mm/s, and 15 mm/s, respectively. For v = 2.2 mm/s, we do not see any flow-induced structure, and the image shows a uniform darkness. For v = 9 mm/s, we observe the formation of a long, bright, extended wake that remains attached to the pin. \textcolor{black}{The bright band-like structure remains stable under flow without showing any significant length variation. However, at the tail end of the stable wake, the presence of some local dynamics may result in weak temporal undulations in force [Fig. S13]. Nonetheless, our turbidity measurements are not sensitive enough to capture such dynamics of the wake.} Interestingly, right after the beginning of the plateau region of the F-v plot, such a bright band starts to show continuous attachment-detachment events with the pin [Fig. 2(d), see movie 2]. The onset velocity of the pin to observe such events correlates with that of the stress fluctuation events in a one-to-one manner. This indicates that the attachment/detachment dynamics of the flow-induced structures are intricately related to the dynamics of steady-state force fluctuations. Interestingly, such sample turbidity under shear (Fig. 2b - 2d) does not depend on the polarization of light. We have tested this by measuring turbidity by placing a linear polarizer in front of the light source and an analyzer in front of the camera. We observe that the sample turbidity for a particular velocity of the pin (in the force-plateau region) remains unchanged for different combinations of the optical axes of the polarizer and the analyzer. Our observations are consistent with the sample turbidity obtained from bulk rheology measurements on the same sample, where no polarization dependence is observed \cite{ghadai2023origin}.
\newline

\begin{figure*}
    \centering
    \includegraphics[height=9.8 cm]{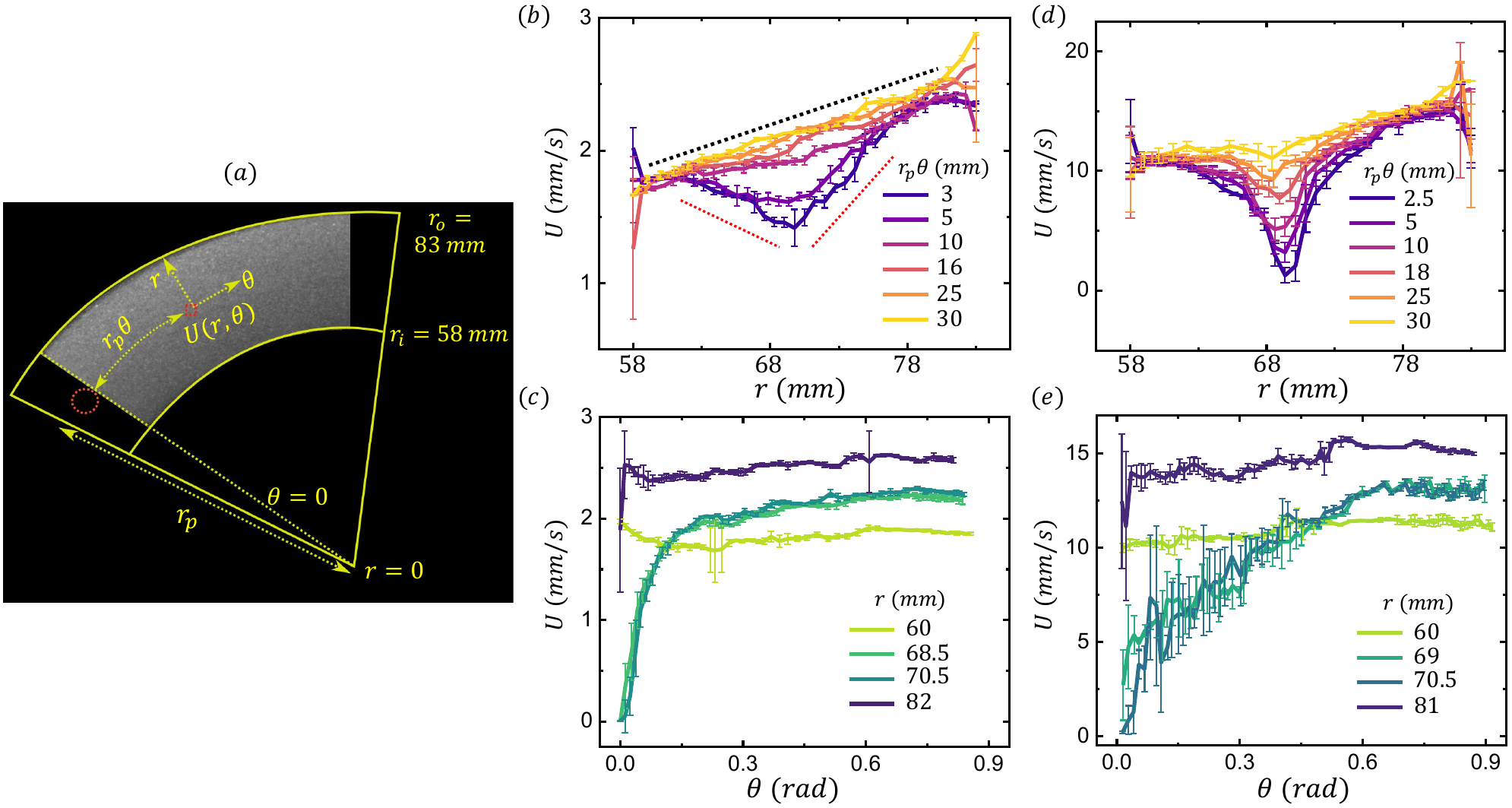}
    \caption{\label{fig:wide} (a) Raw image of the sample-air interface in the sample cell seeded with polystyrene tracer particles for 2wt\% CTAT + 100 mM NaCl system. The red dotted circle is represented as the pin position. (b) and (c) are the radial and azimuthal variation in velocity respectively for a driving velocity of 2 mm/s at $25\,^{\circ}C$. Similarly, (d) and (e) are the radial and azimuthal variation in velocity respectively for a driving velocity of 12 mm/s at room temperature of $25\,^{\circ} C$. The azimuthal ($r_p \theta$) and radial positions ($r$) are indicated by different line colours in the figures.  }
\end{figure*}

Next, to establish a correlation between the attachment detachment dynamics with the observed force fluctuations, we map out the velocity distribution in the system using the particle imaging velocimetry (PIV) technique. As the micellar solution is transparent, 1.5 wt\% of polystyrene tracer particles of size varying between 20-50 $\mu m$ are added to the system to generate the contrast required for PIV. We confirm that adding such a small amount of tracer particles does not modify the rheological properties of the WLM system under study [Fig. S1]. For visualization, we image the top surface of the sample with the pin held fixed and the bottom geometry rotated [Fig. S5]. The recorded images are analyzed using the PIV Lab algorithm using MATLAB software\cite{Thielicke_2021}. Fig. 3(a) shows the typical snapshot of the raw image of the sample-air interface seeded with tracer particles. Here we note that due to a difference between the radial distance of the two walls of the annular geometry from the axis of rotation, even without the presence of the pin the azimuthal velocity increases linearly from the inner to the outer wall.  Fig. 3(b) shows the spatial distribution of velocity in the steady state for v =  2.5 mm/s. The velocity distribution in the presence of the pin (depicted as a black circle in the velocity distribution map) is almost similar to that without the pin. This indicates that for low velocities well below the onset of the plateau of the $F$ vs $v$ curve, the pin perturbs the system only locally, as also seen by a small low-velocity region around the pin [Fig. 3(b)]. When the applied velocity is increased to 12 mm/s (just below the onset of the plateau shown in Fig. 3(c)), a wake having a long, tail-like stationary structure is formed that remains connected to the probe and has much smaller velocity magnitude compared to the other portion of the sample. Interestingly, in the case of Newtonian fluids like glycerol, we observe that the presence of the pin perturbs the velocity distribution of the system only locally even if the velocity is varied over a wide range [Fig. S18]. This marks the striking difference between the flow fields around the pin for WLM system as compared to a high-viscosity Newtonian liquid. As mentioned earlier, we obtain large temporal fluctuations of F for an applied velocity that lies in the early plateau region of the $F-v$ curve where the features of the fluctuations are saw-tooth type. In Fig. 3(d), we show a zoomed-in version of a typical fluctuation event for v = 15 mm/s showing a slow build-up followed by a sharp drop in force.  As indicated in the figure, we mark different regions in the time series as i, ii, iii, and iv. The corresponding spatial distributions of velocity are shown in Fig. 3 (panels i - iv). Just before the sharp force drop (region i), the spatial distribution of velocity shows the formation of a flow-induced long wake-like structure of a low-velocity band behind the pin (similar to the stable wake obtained for v = 12 mm/s). Such a stable wake quickly detaches from the pin and starts to move away from the pin with a much higher velocity compared to the velocity of the moving sample cell. At this point, force also starts to drop. Subsequently, the detached wake disturbs the sample around and pushes until a maximum speed of the sample is reached (iii and iv) and $F$ also reaches a minimum value. After this, the formation of a stable wake connected with the pin again takes place. The gradual increase in $F$ takes place correlating with the gradual elongation of the wake-like structure until the next force-drop event sets in. Such events continue as long as the sample cell velocity is maintained. Our results establish that the force fluctuation events and the formation and detachment of flow-induced structures are correlated in a one-to-one manner. 
\newline

In the increasing branch of the saw-tooth waveform for $F(t)$ [Fig. 4(a)], we find that the length of the wake attached to the pin also increases with time till the maximum value of the force is reached (before the sharp force drop). Here, we define the length of the wake as the length of the region along the direction of azimuthal velocity over which the flow field gets altered due to the presence of the pin. Physically, this length scale signifies the region of influence of the pin for a given driving velocity and is obtained from the PIV measurement [panel i and ii in Fig. 4(a)]. We denote the maximum length of the wake formed within a saw-tooth waveform as $L_{W}$. We further define a rise-time $\Delta t_r$ as the time required for $F(t)$ to rise monotonically from a local minimum value to reach a maximum before it starts to drop abruptly as shown in Fig.  4(a). Earlier studies indicate that changing the degree of screening by varying the salt concentrations can tune the flow-concentration coupling in CTAT WLM\cite{ganapathy2006intermittency}. We observe that a small variation of salt concentration (30 mM to 100 mM NaCl) does not change the nature of the $F-v$ curve except for slight shifting of the plateau onset [Fig. S10]. We also find a similar saw-tooth wave pattern close to the plateau onset for different salt concentrations. We construct a length scale from the force rise time: $L_{t_r} = v \Delta t_r$, where $v$ is the set velocity of the annular geometry at the location of the pin. We observe that $L_{W}$ (obtained from PIV measurement) and $L_{t_r}$ (obtained from the saw tooth behaviour of time-varying force) are very close to each other and vary in a correlated fashion as shown in Fig. 4(b) for a range of salt concentrations.
 Interestingly, we observe from Fig. 4(b) that the length of the wake before detachment increases with increasing salt concentration. Our results highlight that the rise of the force on the probe originates from the stretching of the micellar structure in the wake and the detachment of the structure from the pin results in the sharp drop in the force. Although close to the plateau region of the $F - v$ curve, the nature of most fluctuation events is quite close to the saw-tooth shape, few of the events are more complex (e.g. the first event shown in Fig. 4(a)) showing local non-monotonicity. In such cases, the wake region shows a partial detachment. We ignore such events in our correlation plot in Fig. 4(b). Interestingly, we observe that such partial detachment events become progressively more frequent as we enter deeper inside the plateau region. This observation rationalizes the complex waveform observed in the deep plateau region giving rise to more turbulent/chaotic flows. Apart from the change in salt concentration, temperature also plays a significant role in controlling the structural and mechanical properties of WLM. We find that the overall nature of the $F-v$ curve and the nature of fluctuations remain similar over a range of temperatures from $20\,^{\circ} C$ to $30\,^{\circ} C$ that we probe. We observe that for our system, the plateau onset shifts towards higher velocity with increasing temperature [Fig. S9]. However, estimating the length of the wake $L_{W}$ using PIV analysis at different temperatures remains outside the scope of the present study due to the technical limitation of controlling the sample temperature in twin-drive mode.
 
Next, we probe the effect of the length of the wake on the magnitude of the sudden force drop. We vary the length of the wake $L_{t_r}$ by changing temperature and salt concentration. We define the normalized force drop as $\Delta F_N = \frac{F_i - F_f}{F_i}$, where $F_i$ is the magnitude of force just before the drop and $F_f$ is force just after the drop (indicated in Fig. 4(a)). Fig. 4(c) and 4(d) show a plot of $\Delta F_N$ as a function of $L_{t_r}$ obtained from the time series for different temperatures and salt concentrations, respectively. Here individual point represents a stress drop event. The scattered nature of points indicates a wide distribution of $\Delta F$ and $L_{t_r}$ even for the same temperature/salt concentration. From Fig 4(c) we observe that for a given temperature, a larger $L_{t_r}$ value also implies an enhanced $\Delta F_N$. However, as the temperature increases, longer wakes become more probable but the $\Delta F_N$ values remain almost similar. On the other hand, we observe that both $L_{t_r}$ and $\Delta F_N$ increase with the increasing salt concentration.  Our results imply that a change in micellar structure and entanglement tune the magnitude of the force fluctuation despite the similar nature of the waveform.

\begin{figure*}
    \centering
    \includegraphics[height=6.6 cm]{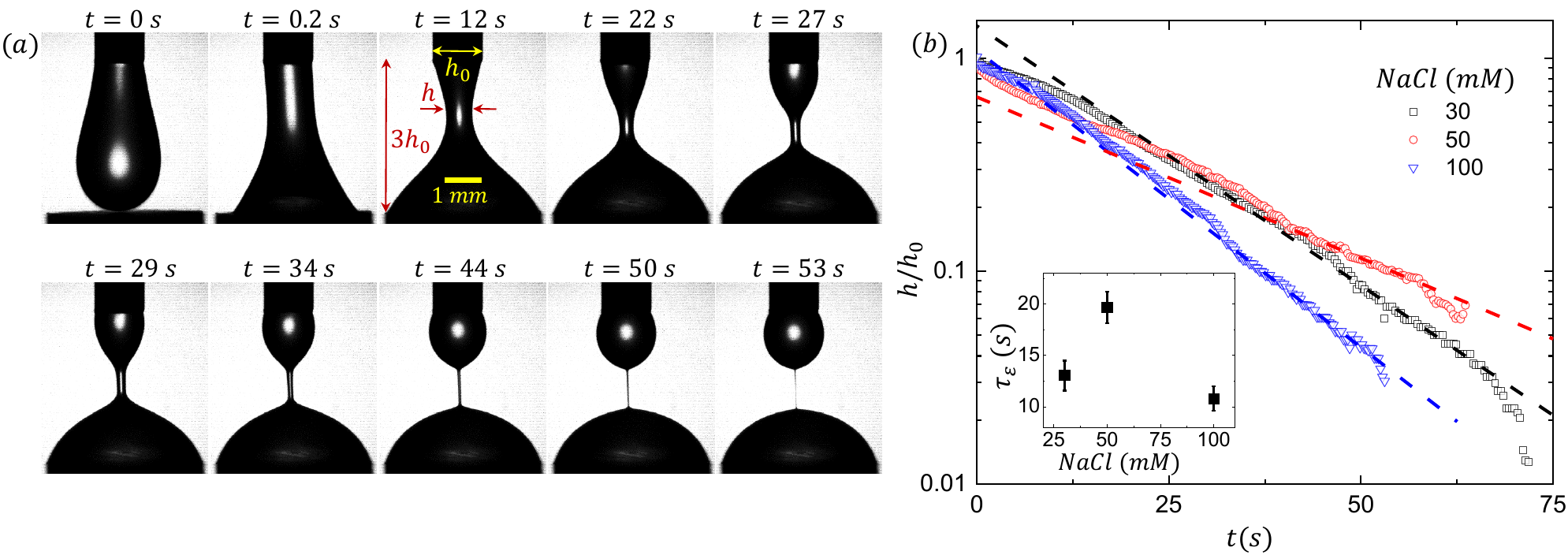}
    \caption{\label{fig:wide} (a) Experimental snapshots at different times for 2wt\% CTAT + 100 mM NaCl system obtained from custom made DoS rheometry to probe the behaviour of the system under extensile deformation. (b) Temporal evolution of the minimum neck diameter of the liquid bridge normalized w.r.t. the needle diameter with varying NaCl concentrations (30mM, 50mM, and 100 mM). The dashed lines are the fit to the data using the elasto-capillary model in order to extract the extensional relaxation times. The average extensional relaxation time of the systems having 30 mM, 50 mM, and 100 mM NaCl are 13s, 19s, and 11s, respectively (inset of (b)). }
\end{figure*}

These results further highlights an intimate connection between the formation and dynamics of the wake in controlling the flow instability of the system. We now study the dynamics of wake formation before the onset of stress fluctuations more quantitatively. Fig. 5(a) shows a typical snapshot of the sample-air interface seeded with polystyrene tracer particles. To quantify the components of the local strain rates we use polar coordinates($r,\theta$) with the origin chosen at the center of rotation of the annular geometry as shown in the schematic in Fig 5(a). Such a choice is consistent with the symmetry of the flow. The radial position of the pin is denoted by $r = r_p$ = 71 mm and $\theta$ = 0. Here, we note that due to the cylindrical shape of the sample cell, the magnitude of the azimuthal velocity increases linearly with the radial position as we go from the inner to the outer boundary of the geometry. We show the radial (along $r$) and azimuthal (along $\theta$) variation in velocity magnitude for 2 mm/s driving velocity in Figs. 5(b) and 5(c) respectively. We observe from Fig. 5(b) that there is a linear variation of the fluid velocity across the shear gap only for a small distance from the pin (given by small $r_p \theta$ values). However, when $r_p \theta$ becomes $\geq$ 10 mm, the presence of the pin is hardly felt. Thus, for lower driving velocities (well below the plateau onset), there is a constant velocity gradient $\frac{\delta U(r, \theta)}{\delta r}$  across the gap. Next, we show the variation of velocity as a function of $\theta$ for different radial positions in Fig. 5(c). We find that only close to the pin location, velocity shows azimuthal variation indicating a non-zero extensile component $\frac{1}{r} \frac{\delta U(r, \theta)}{\delta \theta}$. Nonetheless, such gradient decays sharply to zero with increasing $\theta$. Our results indicate the localized nature of the shear and extensile gradients for low driving velocities. Similar results are also obtained for a Newtonian fluid glycerol [Fig. S19]. Interestingly, the situation changes drastically for higher velocities close to the plateau region as shown in Fig. 5(d) and 5(e) for 12 mm/s driving velocity. We observe a highly non-linear variation of velocity along the radial direction, with a strong gradient close to the radial position of the pin [Fig. 5(d)]. Remarkably, such gradient is observed even for the large values of $r_p \theta$ spanning almost the entire field of view. We also find from Fig. 5(e) that the velocity varies over a large range of $\theta$ in the azimuthal direction near the radial position of the pin. Interestingly, such variation is approximately linear, implying a constant extensile gradient. Close to the plateau regime, the velocity gradient shows strong localization along the radial direction and extended nature along the azimuthal direction. We also find that the radial velocity at the pin position approaches zero (within experimental error, Fig. 5(e)). This indicates that wall slippage is negligible in our case. This implies the extensile nature of the flow and micelles undergo significant strain in this azimuthally extended regime. Interestingly, the long wake formed before the detachment (discussed in Fig. 3 and Fig. 4) coincides with this azimuthally extended region of high-strain deformation that deciphers the formation of a long wake in the system.
\newline

Using the shear and extension rates just before the onset of the unstable regime, we want to now understand the nature of the observed flow instability. For this, we next quantify shear and extensional relaxation time for our system.  We obtain the shear relaxation time from the crossover of the storage ($G'$) and loss ($G''$) moduli by performing a frequency sweep measurement using a Taylor-Couette geometry [Fig. S20]. We obtain a shear relaxation time $\tau_R$ $\sim$ 1 s for 2 wt. \%  CTAT + 100 mM NaCl system at a temperature of $25\,^{\circ}C$. 

The extensional deformation response of the system at room temperature of $25\,^{\circ} C$ is probed using a custom made dripping on to substrate (DoS) rheometry technique following the standard protocol mentioned in the literature\cite{dinic2015extensional,dinic2017pinch,omidvar2019detecting}. We observe the time evolution of a liquid bridge between a modified substrate and the needle that pinches off due to the role of strong capillary force. Fig. 6(a) shows the snapshots of the liquid bridge at different times for 2wt\% CTAT + 100 mM based wormlike micellar system at room temperature of $25\,^{\circ} C$. We record the time evolution of the minimum neck diameter($h$) of the unstable liquid bridge normalized w.r.t. the needle diameter ($h_0$) and extract the extensional relaxation time $\tau_E$ by fitting an exponential curve ($\frac{h}{h_0} \sim e^{\frac{-t}{3\tau_E}}$) as predicted by the elasto-capillary model\cite{entov1997effect}. Fig. 6(b) shows the temporal evolution of the normalized mid-filament diameter for 2wt\% CTAT system with variation of NaCl concentrations (see Fig. S21 for snapshots in the case of 30 mM and 50 mM NaCl) and the dashed lines are the fits using elasto-capillary model. The data is reproducible over multiple set of experiments [Fig. S22]. The extensional relaxation time obtained from DoS fittings over 4 repeated measurements is plotted w.r.t. the variation of NaCl concentrations (see inset of Fig.6(b)).

 \textcolor{black}{Interestingly, 2 wt \% CTAT + 100 mM NaCl does not show shear banding (before unstable/time-varying flows, the velocity profile remains linear). Such a non-banding nature of flow is also expected from the significant slope of stress plateau ($n \sim$ 0.3) in the flow curve and is also confirmed by direct imaging and PIV analysis \cite{ghadai2023origin}.} At high $Wi$, the system directly enters into an unstable flow regime with random time-varying changes in the flow profile. However, in our pin-driving set-up, the force-velocity curve shows a very flat plateau, indicating the possibility of shear-banding. Indeed, our PIV measurement clearly shows that significant non-linearity in the flow profile develops near the plateau onset at much lower driving velocities beyond which the time-varying flow develops. Such a non-linear flow profile originates from the stable wake formation. The estimated value of Reynolds number ($Re = \frac{\rho v d}{\eta}$, where $\rho$ is the density, $v$ is the driving velocity, $d$ is the pin diameter, $\eta$ is the zero shear viscosity) at the plateau onset is much smaller than 1 ($Re \sim 0.005$) signifying that the elasticity plays a major role. \textcolor{black}{The extensile Deborah number \cite{moss2010flow, mohammadigoushki2016sedimentation} ($De_{ext} = \dot{\epsilon}\tau_R$, where $\dot{\epsilon}$ is the maximum extension rate obtained from PIV measurement) at the plateau onset is $\sim$ 1. 
 We further note that the shear ($Wi_S = \dot{\gamma} \tau_R$ ) and extensional ($Wi_E = \dot{\epsilon} \tau_E $) Weissenberg numbers ($\dot{\gamma}$ and $\dot{\epsilon}$ respectively denote the maximum shear and extension rates close to the pin) at the onset of time-varying flow in pin-driving set-up are also close to unity ($Wi_S $, $Wi_E $ = 3). This onset $Wi$ observed for the pin driving set up is significantly lower than the onset $Wi$  of unstable flow in the Taylor-Couette geometry for the same system ($Wi = 16$) \cite{ghadai2023origin}. Nonetheless, in both cases, the instability is driven by fluid elasticity as $Wi >> Re$. We note that $Wi$ alone may not be sufficient to characterize the onset of instabilities, as $Wi$ depends on flow geometries and streamline curvature. We also perform a Pakdel-McKinley analysis for our system to compare the onset of elastic instability \cite{mckinley1996rheological,pakdel1996elastic} (see supplementary information for the calculations). We observe that for Taylor-Couette geometry, such analysis yields an onset value very close to our experimental observation, considering an upper-convected Maxwell model for the system. However, for the pin driving setup, the relevant geometry is the flow past a cylinder. Such an analysis is considerably more complicated, as one needs to estimate additional variables either from numerical modeling or by changing the parameters of the flow geometry. Furthermore, the baseline curvature of the streamlines that come from the circular shape of the sample cell can add further complexities. Exploring these directions remains outside the scope of the present study, but constitutes an important future direction. We believe that the main reason behind the difference in the nature of the flow curves obtained in pin driving and Taylor-Couette set up stems from the fact that in Taylor-Couette geometry, the flow is a simple shear flow to a good approximation. On the other hand, in pin driving, the flow has both shear and extensional components. The coupling between these different modes of deformations through the complex micellar structure can also play a role.} 

\section{Conclusions}
In summary, we investigate the motion of a probe driven through a wormlike micellar fluid at a controlled velocity using a home-built shear cell coupled with in-situ flow visualization. For the first time, we correlate temporal stress fluctuations on the probe and the local shear and extensional deformations in a one-to-one fashion. We find that the flat force plateau obtained in the present setup differs from the finite slope of the stress plateau obtained in bulk rheology measurements for the same WLM system. \textcolor{black}{We find a distinct saw-tooth feature of fluctuations at the onset of the instabilities. Similar features are not observed during bulk rheological measurements of the same system. We also observe some differences in the nature of the flow curve and the onset $Wi$ for stress fluctuations in the pin-driving set up as compared to those obtained for a Taylor-Couette geometry. Such differences can arise from the fact that both shear and extensional deformations play an important role in pin-driving as opposed to the simple shear flow in standard rheological geometries like Taylor-Couette.  Interestingly, the context of flow past a cylinder is geometrically quite similar to the present setup, with an additional complexity of curvature of the sample cell in the pin-driving geometry.} There have been detailed studies of flow instability for both Newtonian, as well as, viscoelastic shear-thinning and Boger fluids  in the context of flow past a cylinder \cite{mckinley1993wake,sheridan1997flow}. For viscoelastic flows, elastic wake instability and the formation of 3-D cellular structure along the length of the cylinder have been observed. \textcolor{black}{Our set up can also be implemented in these cases to not only incorporate flow visualization but also to measure the mechanical force/stress reliably with good resolutions. However, in the current scenario, the pin completes one full rotation in less than 1 s for the pin velocity greater than 450 mm/s. For our WLM system, this sets the upper limit of the velocity as the shear relaxation time is $\sim$ 1 s. If the pin is driven with a higher velocity, the pin retraces its path before the sample fully relaxes. The force response will be dominated by the memory effect in the system, which hinders reliable mechanical measurements at such high driving velocities. Even at such high velocities, the Reynolds number just crosses 1 for Newtonian fluid glycerol. Thus,  we need to go to a much higher velocity to induce inertia-driven instability in the case of glycerol. The reliability of force measurement and the high frame rate requirement need further optimization of the set up. Nonetheless, these directions will be very interesting to explore in the context of the motion of small probes in different viscoelastic systems.}

Our study reveals the formation of flow-induced stable wakes just before the onset of force fluctuations. At higher driving velocities, the attachment detachment dynamics of the shear-induced micellar structures in the wake region with the probe remains strongly correlated with the stress fluctuations in the system. In the wake region, we get flow-induced turbidity which is visible even under the ambient light. This turbid band forms due to the scattering of light from the shear-induced structures in the non-linear flow regime. Interestingly, the scattered intensity is not sensitive to the polarization of light which we verify using different orientations of optical axes of a linear polarizer-analyzer pair. Understanding the microscopic origin of the sample turbidity will be an interesting future research direction. Careful measurement of sample birefringence might give us important insights. 
\textcolor{black}{In the present study, we probe the stress fluctuations under a controlled velocity. Our preliminary data on controlled stress measurements reproduce the damped harmonic oscillations and bursts observed in the ball drop experiments in the CTAT-NaCl systems\cite{kumar2012oscillatory}. Hence, stress/force-controlled measurements can yield significantly different temporal features of fluctuations in comparison to rate/velocity-controlled measurements.} Our experiments can probe burst statistics much better as the conventional ball drop experiments are limited by the finite tube length. However, we must remember that the probe is held rigidly in the current setup which does not allow any rotational or lateral motion of the probe. However, in the case of ball-drop experiments, no such restrictions are present. This can affect the dynamics. Introducing such flexibility in our set-up will be an interesting direction for future studies.

\section*{Supplementary material}
   See the supplementary material for information about different imaging set-ups, data analysis and additional measurements.
\section*{Acknowledgements}
    S.M. acknowledges the SERB (Under DST, Government of India) for Ramanujan Fellowship. We thank Nitin Kumar, Sandra Lerouge, Chandi Sasmal, Naveen Kumar Chandra, Vaibhav Parmar  and Harikrishna Sahu for useful discussions and feedback. We also acknowledge the detailed and constructive comments by the reviewers towards improving the manuscript.   
\section*{Author Declarations}
\subsection*{Conflict of Interest}
The authors have no conflicts to disclose.
\subsection{Author Contributions}
A.G. and P.K.B. performed the experiments. A.G. set up the in-situ imaging and DoS rheometry. A.G. and S.M. analyzed the data and wrote the manuscript.
\section*{Data Availability}
The data that support the findings of this study are available from the corresponding author upon reasonable request.
\bibliography{Ref}

@article{stokes1851effect,
  title={On the effect of the internal friction of fluids on the motion of pendulums},
  author={Stokes, George Gabriel and others},
  year={1851},
  publisher={Pitt Press Cambridge}
}

@article{li2020rheological,
  title={Rheological behavior of a wormlike micelle and an amphiphilic polymer combination for enhanced oil recovery},
  author={Li, Xinxin and Sarsenbekuly, Bauyrzhan and Yang, Hongbin and Huang, Zitong and Jiang, Haizhuang and Kang, Xin and Li, Menglan and Kang, Wanli and Luo, Peng},
  journal={Physics of Fluids},
  volume={32},
  number={7},
  year={2020},
  publisher={AIP Publishing}
}

@article{bewersdorff1988behaviour,
  title={The behaviour of drag-reducing cationic surfactant solutions},
  author={Bewersdorff, H -W and Ohlendorf, D},
  journal={Colloid and Polymer Science},
  volume={266},
  pages={941--953},
  year={1988},
  publisher={Springer}
}

@incollection{ezrahi2007daily,
  title={Daily applications of systems with wormlike micelles},
  author={Ezrahi, Shmaryahu and Tuval, Eran and Aserin, Abraham and Garti, Nissim},
  booktitle={Giant Micelles},
  pages={515--544},
  year={2007},
  publisher={CRC Press}
}

@book{dreiss2017wormlike,
  title={Wormlike micelles: advances in systems, characterisation and applications},
  author={Dreiss, C{\'e}cile A and Feng, Yujun},
  year={2017},
  publisher={Royal Society of Chemistry}
}

@article{cates1987reptation,
  title={Reptation of living polymers: dynamics of entangled polymers in the presence of reversible chain-scission reactions},
  author={Cates, ME},
  journal={Macromolecules},
  volume={20},
  number={9},
  pages={2289--2296},
  year={1987},
  publisher={ACS Publications}
}

@article{salmon2003velocity,
  title={Velocity profiles in shear-banding wormlike micelles},
  author={Salmon, Jean-Baptiste and Colin, Annie and Manneville, S{\'e}bastien and Molino, Fran{\c{c}}ois},
  journal={Physical review letters},
  volume={90},
  number={22},
  pages={228303},
  year={2003},
  publisher={APS}
}

@article{helgeson2009relating,
  title={Relating shear banding, structure, and phase behavior in wormlike micellar solutions},
  author={Helgeson, Matthew E and Reichert, Matthew D and Hu, Y Thomas and Wagner, Norman J},
  journal={Soft Matter},
  volume={5},
  number={20},
  pages={3858--3869},
  year={2009},
  publisher={Royal Society of Chemistry}
}

@article{mohammadigoushki2016flow,
  title={A flow visualization and superposition rheology study of shear-banding wormlike micelle solutions},
  author={Mohammadigoushki, Hadi and Muller, Susan J},
  journal={Soft matter},
  volume={12},
  number={4},
  pages={1051--1061},
  year={2016},
  publisher={Royal Society of Chemistry}
}

@article{fardin2010elastic,
  title={Elastic turbulence in shear banding wormlike micelles},
  author={Fardin, Marc-Antoine and Lopez, Diego and Croso, J and Gr{\'e}goire, Guillaume and Cardoso, Olivier and McKinley, Garett Huw and Lerouge, Sandra},
  journal={Physical review letters},
  volume={104},
  number={17},
  pages={178303},
  year={2010},
  publisher={APS}
}

@article{ghadai2023origin,
  title={Origin of steady state stress fluctuations in a shear-thinning worm-like micellar system},
  author={Ghadai, Abhishek and Bera, Pradip Kumar and Majumdar, Sayantan},
  journal={Physics of Fluids},
  volume={35},
  number={6},
  year={2023},
  publisher={AIP Publishing}
}

@article{majumdar2011universality,
  title={Universality and scaling behavior of injected power in elastic turbulence in wormlike micellar gel},
  author={Majumdar, Sayantan and Sood, AK},
  journal={Physical Review E—Statistical, Nonlinear, and Soft Matter Physics},
  volume={84},
  number={1},
  pages={015302},
  year={2011},
  publisher={APS}
}

@article{bandyopadhyay2000observation,
  title={Observation of chaotic dynamics in dilute sheared aqueous solutions of CTAT},
  author={Bandyopadhyay, Ranjini and Basappa, Geetha and Sood, AK},
  journal={Physical Review Letters},
  volume={84},
  number={9},
  pages={2022},
  year={2000},
  publisher={APS}
}

@article{ganapathy2006intermittency,
  title={Intermittency route to rheochaos in wormlike micelles with flow-concentration coupling},
  author={Ganapathy, Rajesh and Sood, AK},
  journal={Physical review letters},
  volume={96},
  number={10},
  pages={108301},
  year={2006},
  publisher={APS}
}

@article{ganapathy2008spatiotemporal,
  title={Spatiotemporal nematodynamics in wormlike micelles en route to rheochaos},
  author={Ganapathy, Rajesh and Majumdar, Sayantan and Sood, AK},
  journal={Physical Review E—Statistical, Nonlinear, and Soft Matter Physics},
  volume={78},
  number={2},
  pages={021504},
  year={2008},
  publisher={APS}
}

@article{mohammadigoushki2017inertio,
  title={Inertio-elastic instability in Taylor-Couette flow of a model wormlike micellar system},
  author={Mohammadigoushki, Hadi and Muller, Susan J},
  journal={Journal of Rheology},
  volume={61},
  number={4},
  pages={683--696},
  year={2017},
  publisher={AIP Publishing}
}

@article{perge2014inertio,
  title={Inertio-elastic instability of non shear-banding wormlike micelles},
  author={Perge, Christophe and Fardin, Marc-Antoine and Manneville, S{\'e}bastien},
  journal={Soft Matter},
  volume={10},
  number={10},
  pages={1450--1454},
  year={2014},
  publisher={Royal Society of Chemistry}
}

@article{jayaraman2003oscillations,
  title={Oscillations of a solid sphere falling through a wormlike micellar fluid},
  author={Jayaraman, Anandhan and Belmonte, Andrew},
  journal={Physical Review E},
  volume={67},
  number={6},
  pages={065301},
  year={2003},
  publisher={APS}
}

@article{mohammadigoushki2016sedimentation,
  title={Sedimentation of a sphere in wormlike micellar fluids},
  author={Mohammadigoushki, Hadi and Muller, Susan J},
  journal={Journal of Rheology},
  volume={60},
  number={4},
  pages={587--601},
  year={2016},
  publisher={AIP Publishing}
}

@article{zhang2018unsteady,
  title={Unsteady sedimentation of a sphere in wormlike micellar fluids},
  author={Zhang, Yiran and Muller, Susan J},
  journal={Physical Review Fluids},
  volume={3},
  number={4},
  pages={043301},
  year={2018},
  publisher={APS}
}

@article{chen2004flow,
  title={Flow of a wormlike micelle solution past a falling sphere},
  author={Chen, Sheng and Rothstein, Jonathan P},
  journal={Journal of Non-Newtonian Fluid Mechanics},
  volume={116},
  number={2-3},
  pages={205--234},
  year={2004},
  publisher={Elsevier}
}

@article{wu2021linear,
  title={Linear versus branched: flow of a wormlike micellar fluid past a falling sphere},
  author={Wu, Shijian and Mohammadigoushki, Hadi},
  journal={Soft Matter},
  volume={17},
  number={16},
  pages={4395--4406},
  year={2021},
  publisher={Royal Society of Chemistry}
}

@article{kumar2012oscillatory,
  title={Oscillatory settling in wormlike-micelle solutions: bursts and a long time scale},
  author={Kumar, Nitin and Majumdar, Sayantan and Sood, Aditya and Govindarajan, Rama and Ramaswamy, Sriram and Sood, AK},
  journal={Soft Matter},
  volume={8},
  number={16},
  pages={4310--4313},
  year={2012},
  publisher={Royal Society of Chemistry}
}

@article{vasquez2007network,
  title={A network scission model for wormlike micellar solutions: I. Model formulation and viscometric flow predictions},
  author={Vasquez, Paula A and McKinley, Gareth H and Cook, L Pamela},
  journal={Journal of non-newtonian fluid mechanics},
  volume={144},
  number={2-3},
  pages={122--139},
  year={2007},
  publisher={Elsevier}
}

@article{sasmal2021unsteady,
  title={Unsteady motion past a sphere translating steadily in wormlike micellar solutions: A numerical analysis},
  author={Sasmal, Chandi},
  journal={Journal of Fluid Mechanics},
  volume={912},
  pages={A52},
  year={2021},
  publisher={Cambridge University Press}
}

@article{sasmal2022effect,
  title={Effect of micelle breaking rate and wall slip on unsteady motion past a sphere translating steadily in wormlike micellar solutions},
  author={Sasmal, Chandi},
  journal={Physics of Fluids},
  volume={34},
  number={7},
  year={2022},
  publisher={AIP Publishing}
}

@article{moss2010flow,
  title={Flow of wormlike micelle solutions through a periodic array of cylinders},
  author={Moss, Geoffrey R and Rothstein, Jonathan P},
  journal={Journal of non-newtonian fluid mechanics},
  volume={165},
  number={1-2},
  pages={1--13},
  year={2010},
  publisher={Elsevier}
}

@article{dey2018viscoelastic,
  title={Viscoelastic fluid-structure interactions between a flexible cylinder and wormlike micelle solution},
  author={Dey, Anita A and Modarres-Sadeghi, Yahya and Rothstein, Jonathan P},
  journal={Physical Review Fluids},
  volume={3},
  number={6},
  pages={063301},
  year={2018},
  publisher={APS}
}

@article{fardin2009taylor,
  title={Taylor-like vortices in shear-banding flow of giant micelles},
  author={Fardin, Marc-Antoine and Lasne, Benoit and Cardoso, Olivier and Gr{\'e}goire, Guillaume and Argentina, M and Decruppe, JP and Lerouge, Sandra},
  journal={Physical review letters},
  volume={103},
  number={2},
  pages={028302},
  year={2009},
  publisher={APS}
}

@article{ganapathy2008spatiotemporal2,
  title={Spatiotemporal dynamics of shear induced bands en route to rheochaos},
  author={Ganapathy, Rajesh and Majumdar, Sayantan and Sood, AK},
  journal={The European Physical Journal B},
  volume={64},
  pages={537--542},
  year={2008},
  publisher={Springer}
}

@article{dinic2015extensional,
  title={Extensional relaxation times of dilute, aqueous polymer solutions},
  author={Dinic, Jelena and Zhang, Yiran and Jimenez, Leidy Nallely and Sharma, Vivek},
  journal={ACS Macro Letters},
  volume={4},
  number={7},
  pages={804--808},
  year={2015},
  publisher={ACS Publications}
}

@article{dinic2017pinch,
  title={Pinch-off dynamics and dripping-onto-substrate (DoS) rheometry of complex fluids},
  author={Dinic, Jelena and Jimenez, Leidy Nallely and Sharma, Vivek},
  journal={Lab on a Chip},
  volume={17},
  number={3},
  pages={460--473},
  year={2017},
  publisher={Royal Society of Chemistry}
}

@article{Thielicke_2021, doi = {10.5334/jors.334}, url = {https://doi.org/10.5334%2Fjors.334}, year = 2021, month = {may}, publisher = {Ubiquity Press, Ltd.}, volume = {9}, number = {1}, pages = {12}, author = {William Thielicke and Ren{\'{e}} Sonntag}, title = {Particle Image Velocimetry for {MATLAB}: Accuracy and enhanced algorithms in {PIVlab}}, journal = {Journal of Open Research Software} }

@article{dhar2019signature,
  title={Signature of jamming under steady shear in dense particulate suspensions},
  author={Dhar, Subhransu and Chattopadhyay, Sebanti and Majumdar, Sayantan},
  journal={Journal of Physics: Condensed Matter},
  volume={32},
  number={12},
  pages={124002},
  year={2019},
  publisher={IOP Publishing}
}

@article{omidvar2019detecting,
  title={Detecting wormlike micellar microstructure using extensional rheology},
  author={Omidvar, Rose and Wu, Shijian and Mohammadigoushki, Hadi},
  journal={Journal of Rheology},
  volume={63},
  number={1},
  pages={33--44},
  year={2019},
  publisher={AIP Publishing}
}

@article{entov1997effect,
  title={Effect of a spectrum of relaxation times on the capillary thinning of a filament of elastic liquid},
  author={Entov, VM and Hinch, EJ},
  journal={Journal of Non-Newtonian Fluid Mechanics},
  volume={72},
  number={1},
  pages={31--53},
  year={1997},
  publisher={Elsevier}
}

@article{calabrese2015rheology,
  title={The rheology and microstructure of branched micelles under shear},
  author={Calabrese, Michelle A and Rogers, Simon A and Murphy, Ryan P and Wagner, Norman J},
  journal={Journal of Rheology},
  volume={59},
  number={5},
  pages={1299--1328},
  year={2015},
  publisher={AIP Publishing}
}

@article{rassolov2022kinetics,
  title={Kinetics of shear banding flow formation in linear and branched wormlike micelles},
  author={Rassolov, Peter and Scigliani, Alfredo and Mohammadigoushki, Hadi},
  journal={Soft Matter},
  volume={18},
  number={32},
  pages={6079--6093},
  year={2022},
  publisher={Royal Society of Chemistry}
}

@article{pasquino2023startup,
  title={On the startup behavior of wormlike micellar networks: The effect of different salts bound to the same surfactant molecule},
  author={Pasquino, Rossana and Avallone, Pietro Renato and Costanzo, Salvatore and Inbal, Ionita and Danino, Dganit and Ianniello, Vincenzo and Ianniruberto, Giovanni and Marrucci, Giuseppe and Grizzuti, Nino},
  journal={Journal of Rheology},
  volume={67},
  number={2},
  pages={353--364},
  year={2023},
  publisher={AIP Publishing}
}

@article{mohammadigoushki2019transient,
  title={Transient evolution of flow profiles in a shear banding wormlike micellar solution: Experimental results and a comparison with the VCM model},
  author={Mohammadigoushki, Hadi and Dalili, Alireza and Zhou, Lin and Cook, Pamela},
  journal={Soft matter},
  volume={15},
  number={27},
  pages={5483--5494},
  year={2019},
  publisher={Royal Society of Chemistry}
}

@article{perge2014surfactant,
  title={Surfactant micelles: Model systems for flow instabilities of complex fluids},
  author={Perge, Christophe and Fardin, Marc-Antoine and Manneville, S{\'e}bastien},
  journal={The European Physical Journal E},
  volume={37},
  pages={1--12},
  year={2014},
  publisher={Springer}
}

@article{divoux2016shear,
  title={Shear banding of complex fluids},
  author={Divoux, Thibaut and Fardin, Marc A and Manneville, Sebastien and Lerouge, Sandra},
  journal={Annual Review of Fluid Mechanics},
  volume={48},
  number={1},
  pages={81--103},
  year={2016},
  publisher={Annual Reviews}
}

@article{mckinley1993wake,
  title={The wake instability in viscoelastic flow past confined circular cylinders},
  author={McKinley, Gareth H and Armstrong, Robert C and Brown, Robert},
  journal={Philosophical Transactions of the Royal Society of London. Series A: Physical and Engineering Sciences},
  volume={344},
  number={1671},
  pages={265--304},
  year={1993},
  publisher={The Royal Society London}
}

@article{sheridan1997flow,
  title={Flow past a cylinder close to a free surface},
  author={Sheridan, John and Lin, J-C and Rockwell, D},
  journal={Journal of Fluid Mechanics},
  volume={330},
  pages={1--30},
  year={1997},
  publisher={Cambridge University Press}
}

@article{mckinley1996rheological,
  title={Rheological and geometric scaling of purely elastic flow instabilities},
  author={McKinley, Gareth H and Pakdel, Peyman and {\"O}ztekin, Alparslan},
  journal={Journal of Non-Newtonian Fluid Mechanics},
  volume={67},
  pages={19--47},
  year={1996},
  publisher={Elsevier}
}

@article{pakdel1996elastic,
  title={Elastic instability and curved streamlines},
  author={Pakdel, Peyman and McKinley, Gareth H},
  journal={Physical Review Letters},
  volume={77},
  number={12},
  pages={2459},
  year={1996},
  publisher={APS}
}


\clearpage

\newpage
\widetext

\begin{center}

\textbf{\Large Supplementary Information: Instability and stress fluctuations of a probe driven through a worm-like micellar fluid}
\end{center}
\setcounter{section}{0}
\setcounter{figure}{0}
\setcounter{page}{1}

\renewcommand{\thefigure}{S\arabic{figure}}
\renewcommand{\thesection}{S\arabic{section}}

\textbf{\large Access to supplementary movies:}

Movie1: \url{https://drive.google.com/file/d/11aRti4bhqeYOLmDUaAcH3_blkVIx1_OB/view?usp=sharing}

Movie2: \url{https://drive.google.com/file/d/1gRcIErIi70lxFFuuaeOf8YpO__VD0NQd/view?usp=sharing}

\section{Movie Descriptions}
\subsection{Movie 1}
   This movie depicts the animation of the set-up where a pin attached to a shaft is moving inside the annular region. 

\subsection{Movie 2}
   This movie depicts the detachment-attachment of wake for 2wt\% CTAT + 100 mM NaCl system at a driving velocity of 15 mm/s obtained from in-situ turbidity measurements (see Fig. S4 for the set-up). The movie is recorded at 100 frames/sec and played in real time. The scale bar is shown by the yellow line and is of 10 mm.

\section{Calculation of $M_{crit}$ for the onset of elastic instabilities}
The following equations are adapted from McKinley \textit{et}. al., Journal of Non-Newtonian Fluid Mech. (1996).

\subsection{Taylor-Couette flow}

For the Taylor-Couette geometry (inner cylinder radius = 10 mm, gap = 1 mm)

$$M_{crit} = \sqrt{2}\Lambda^{1/2} Wi $$

where, $M_{crit}$ is the critical magnitude for the onset of elastic instability, $\Lambda$ is the aspect ratio (ratio of gap and inner cylinder radius) and $Wi$ is the Weissenberg number.

For 2wt\% CTAT + 100 mM NaCl wormlike micellar system, the estimated value of $M_{crit}$ at the onset elastic turbulence ($Wi = 16$) is 7.15. This value is quite close to the value that is reported using numerical linear stability calculation for upper convected Maxwell model.

\subsection{Pin driving }

If we assume that the pin driving set up is identical to flow past a cylinder (probe radius = 2.5 mm, width of the sample cell = 25 mm )

$$M_{crit} = De\, [a\sqrt{2} \,+\, \frac{b\sqrt{2}}{\Lambda} ]$$
where, $De$ is Deborah number, $\Lambda$ is the aspect ratio (ratio of half width of the sample cell and probe radius), both $a$ and $b$ are numerical constants that depend on the curvature of the streamlines. 

The reported value of $M_{crit}$ is 6.08 obtained from linear regression and numerical calculations. 
For a = 1, we get the value of b = 48. 

 The accurate estimation of both a and b is currently out of scope of the present study as it requires to compute numerically or change the parameters of the flow geometry.

\section{Supplementary Figures}  

\begin{figure}[hbt!]
\centering
\includegraphics[height=6.5cm]{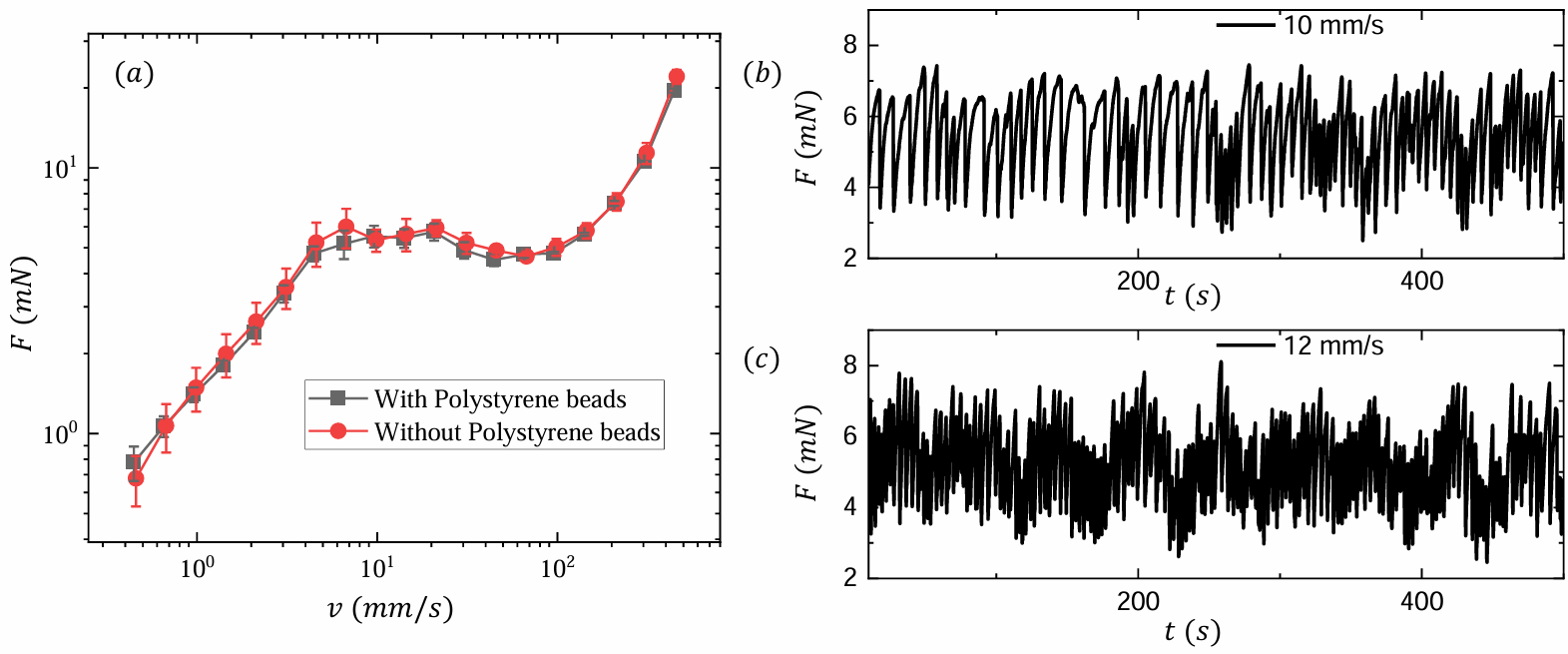}
\caption{\label{fig:wide} (a) Comparison of $F-v$ curve for 2wt\% CTAT + 100 mM NaCl with embeded polystyrene beads (black squares) and without added tracers (red circles) at a temperature of $25^{\circ} C$ for pin diameter of 5 mm. Both the curves overlap on each other. Temporal fluctuations of force at driving velocities (mentioned in the inset) corresponding to the plateau region show saw-tooth feature of fluctuations in the case of micellar system embeded with tracer beads (panel (b) and (c)).   }
\end{figure}

\begin{figure}[hbt!]
\centering
\includegraphics[height=8cm]{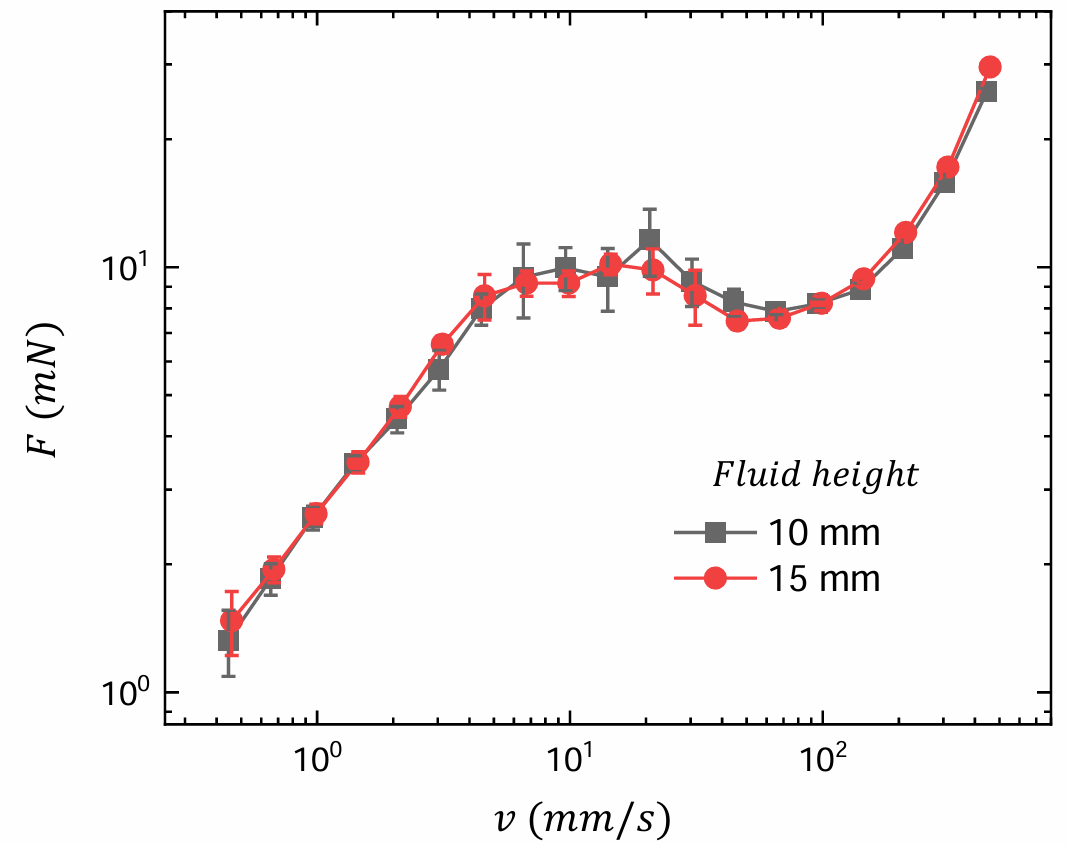}
\caption{\label{fig:wide} $F-v$ curve for 2wt\% CTAT + 100 mM NaCl system at $25^{\circ}C$ for fluid height of 10 mm (black squares) and 15 mm (red circles). Error bars are calculated from the standard deviation of 4 repeated runs. }
\end{figure}

\begin{figure}
\centering
\includegraphics[height=7cm]{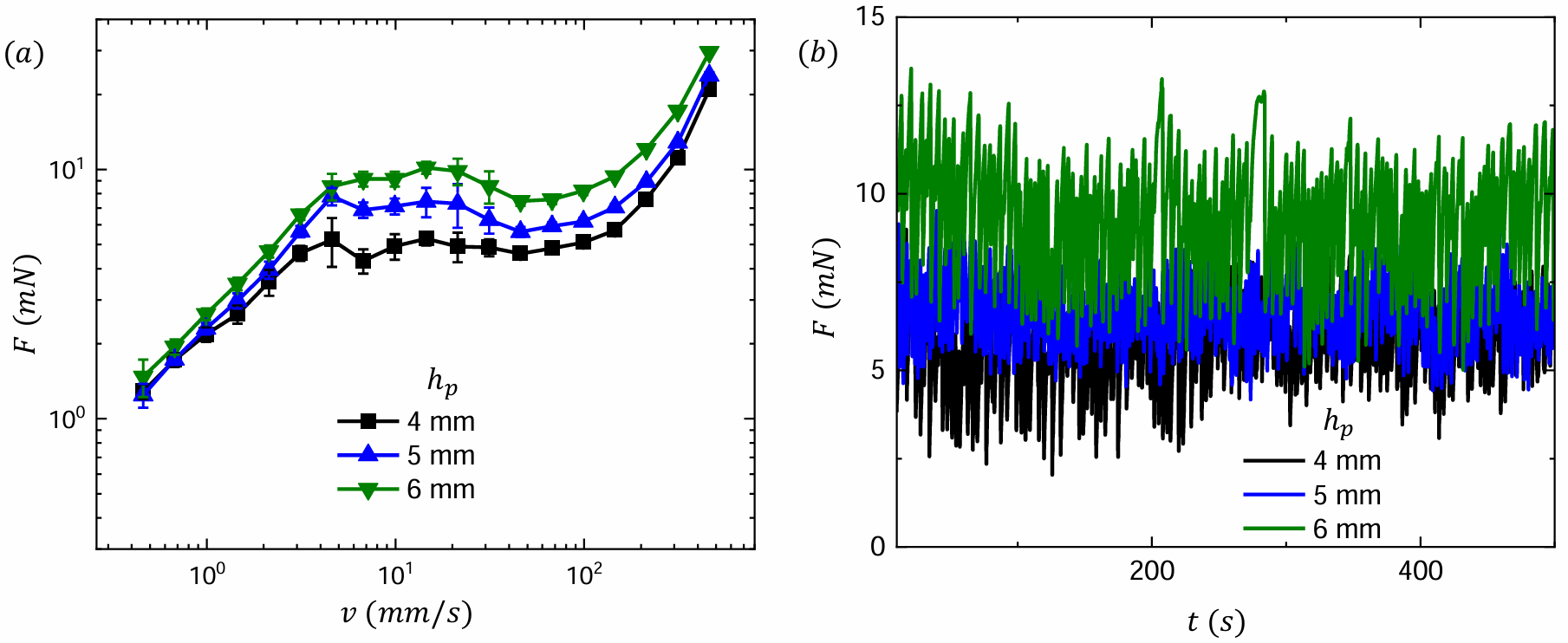}
\caption{\label{fig:wide} (a) Force($F$) - velocity($v$) curve for 2wt\% CTAT + 100 mM NaCl micellar system at $25^{\circ}C$ with change in the length of the submerged portion ($h_p$) of the pin having 5 mm thickness. $F-v$ curve upshifts when $h_p$ is increased. (b) Temporal force fluctuations at a constant driving velocity of 10 mm/s that corresponds to the plateau region for different values of $h_p$. The saw-tooth feature is evident in all the case. However, there is a systematic upshift of the mean force when $h_p$ is increased. }
\end{figure}
    
\begin{figure}
\centering
\includegraphics[height=7.2cm]{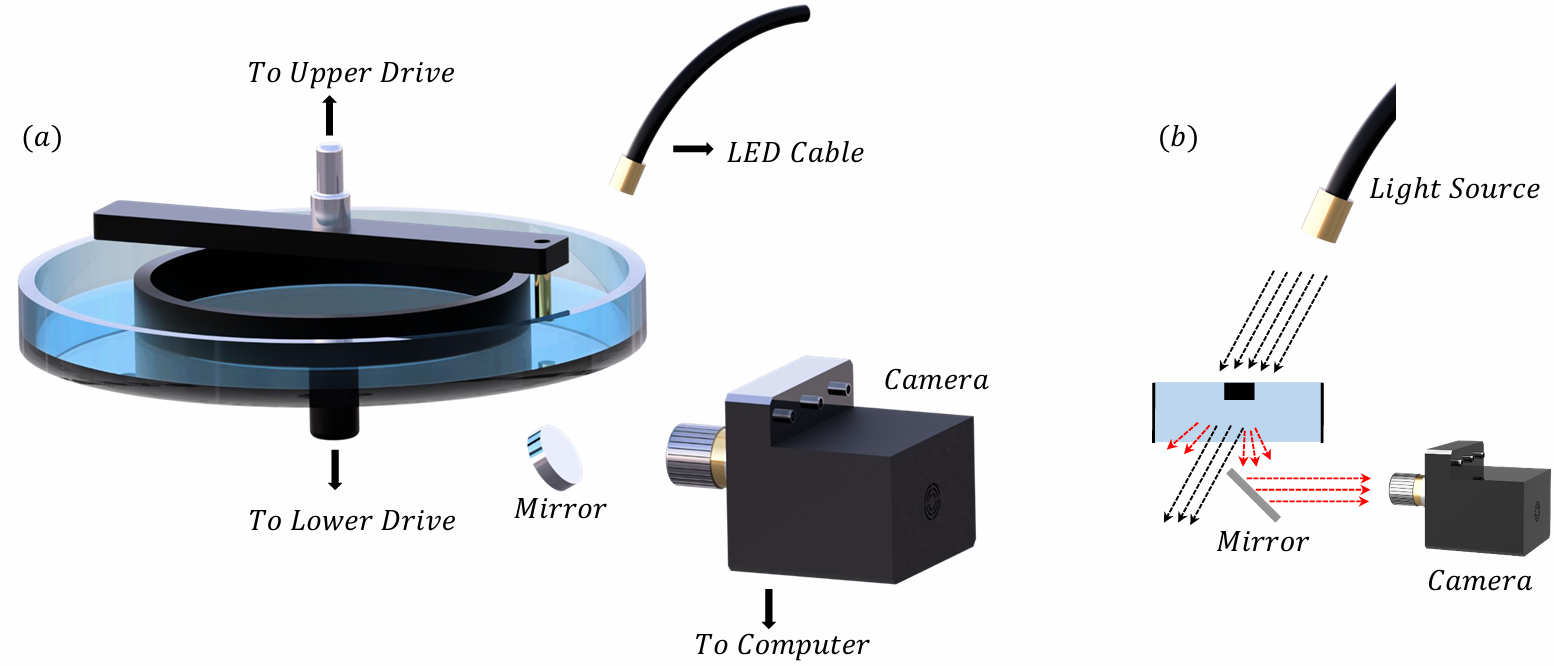}
\caption{\label{fig:wide} (a) Schematic of the imaging set-up for in-situ turbidity measurements. The sample cell (transparent) containing the micellar solution is shined with white light using a LED cable. The scattered light in transmission mode is collected with the help of a fast camera (Phantom Miro C210) fitted with a TV lens (Pentax) by placing a mirror at $45^o$ below the sample cell as shown. (b) Ray diagram of the set up showing the LED light is kept off-axis of the camera to capture the scattering from the turbid structures. The sample appears dark when no flow induced structure is present and presence of any turbid structure will scatter light and will show enhanced brightness.   }
\end{figure}

\begin{figure}
\centering
\includegraphics[height=9cm]{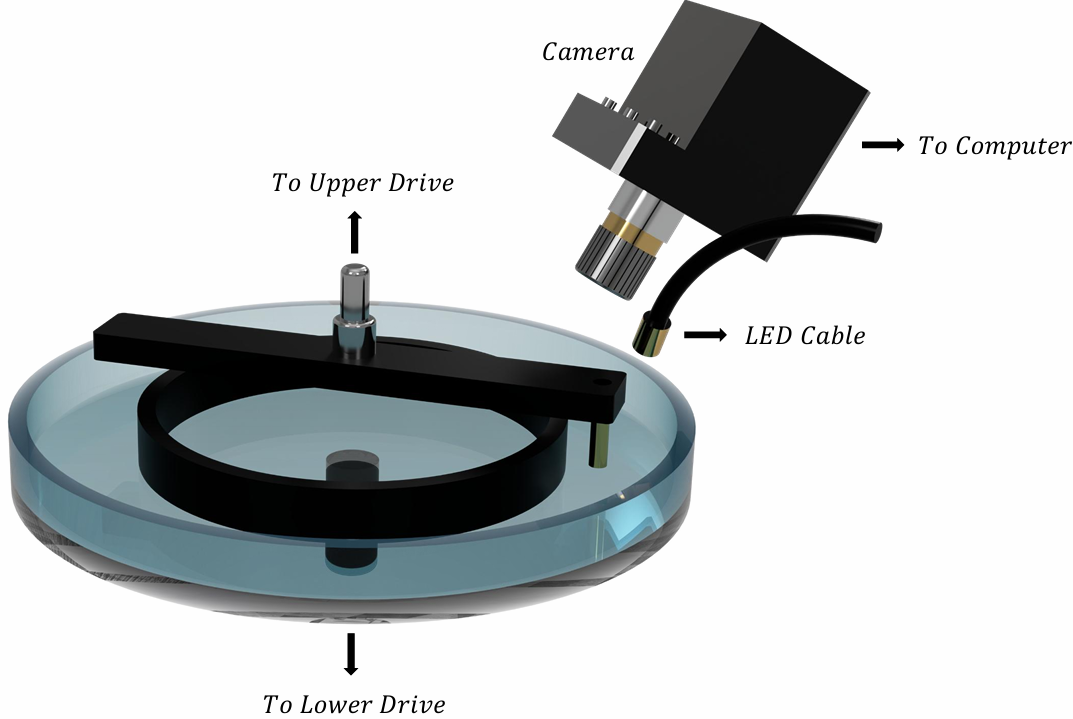}
\caption{\label{fig:wide} Schematic of the imaging set-up to probe the flow field near the pin at sample-air interface. The sample cell containing the micellar solution embeded with polystyrene tracer particles is shined with white light using a LED cable. The reflected light is collected with the help of a fast camera (Phantom Miro C210) fitted with a TV lens (Pentax). }
\end{figure}

\begin{figure}
\centering
\includegraphics[height=8cm]{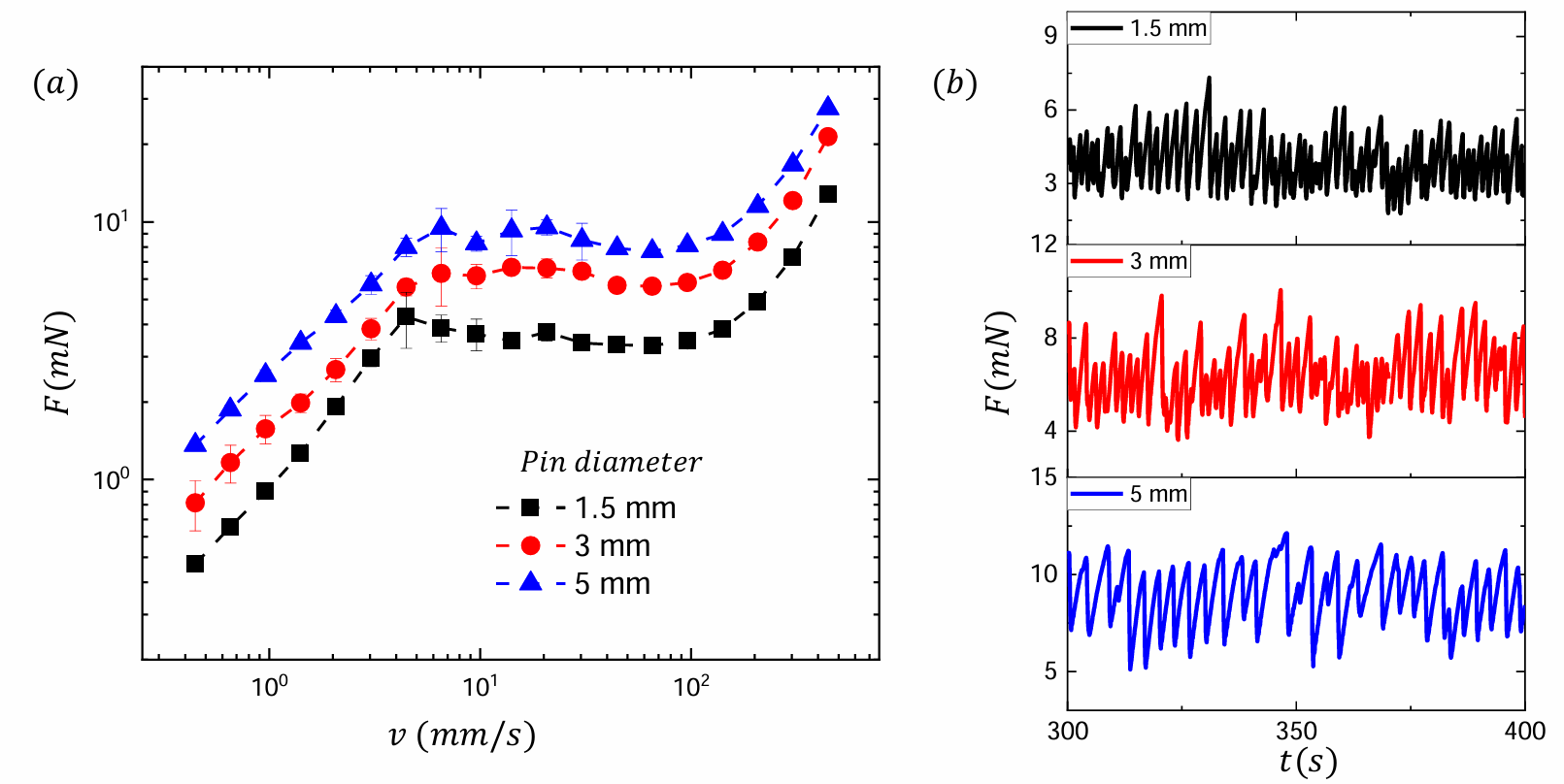}
\caption{\label{fig:wide} (a) Force($F$) - velocity($v$) curve for 2wt\% CTAT + 100 mM NaCl micellar system at $25^{\circ}C$ with variation of pin diameters. $F-v$ curve upshifts when the pin diameter is increased. (b) Temporal force fluctuations at a constant driving velocity of 10 mm/s that corresponds to the plateau region for different pins (black for 1.5 mm, red for 3 mm and blue for 5 mm pin diameter). The saw-tooth feature is evident in all the case. }
\end{figure}

\begin{figure}
\centering
\includegraphics[height=10cm]{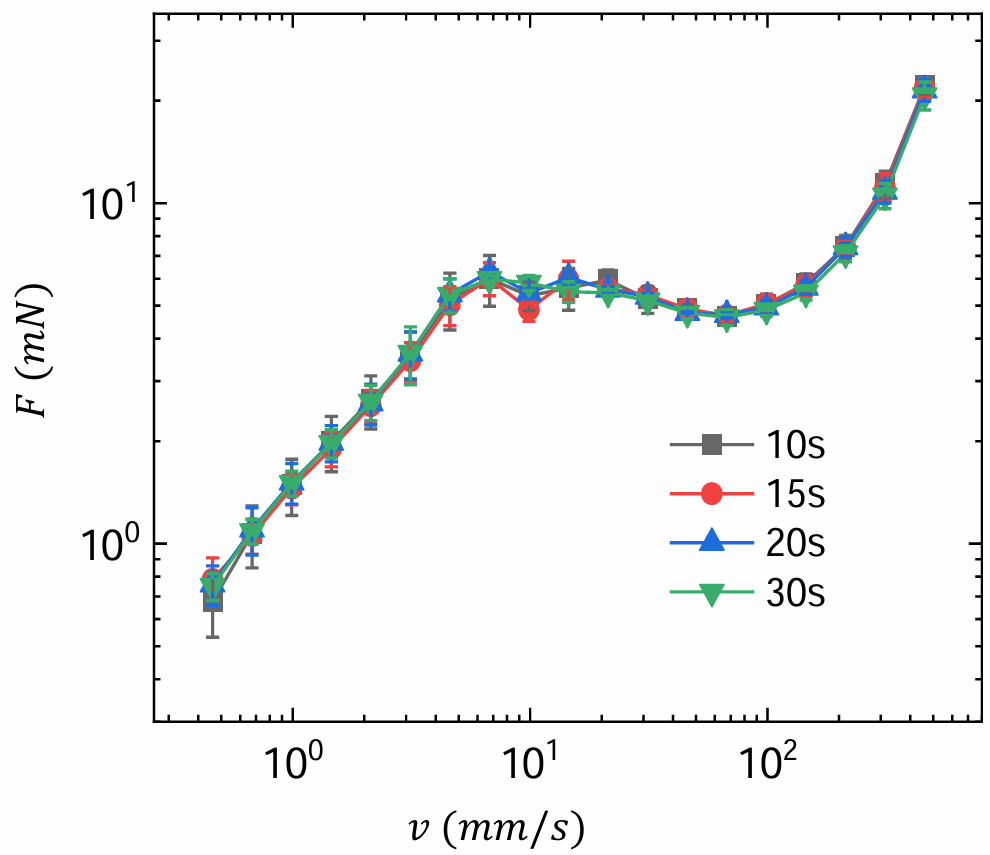}
\caption{\label{fig:wide} $F-v$ curve for 2wt\% CTAT + 100 mM NaCl micellar system at 25℃ with variation of data sampling/acquisition time. The nature of the $F-v$ does not change with change of waiting time.  }
\end{figure}

\begin{figure}
\centering
\includegraphics[height=9cm]{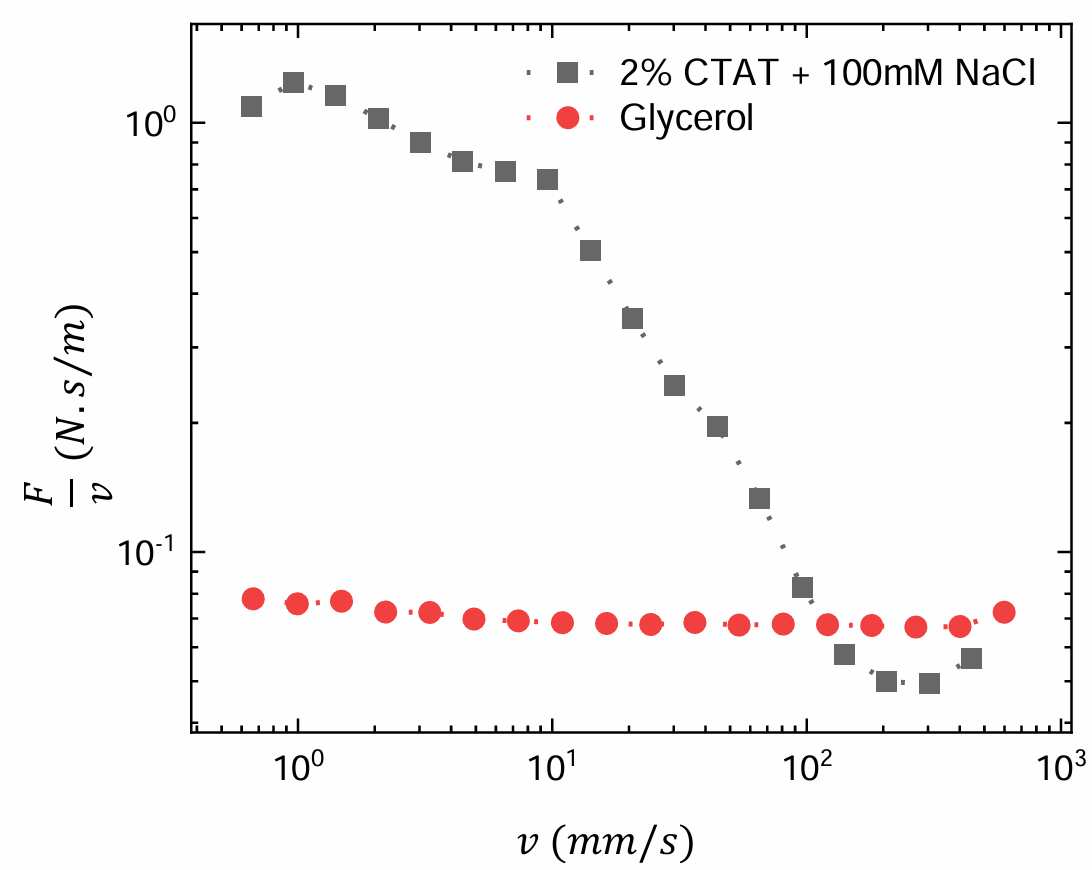}
\caption{\label{fig:wide}  $F⁄v$ versus $v$ curve for 2wt\% CTAT + 100 mM NaCl (black squares) and glycerol (red circles) respectively at $25^{\circ}C$. $F/v$ shows strong decrease beyond the plateau onset ($v \sim$ 10 mm/s). The value of $F⁄v$ is smaller in the case of the micellar solution in compared to glycerol beyond the force plateau (velocities higher than 100 mm/s) in the $F-v$ curve of the micellar system.  }
\end{figure}

\begin{figure}
\centering
\includegraphics[height=7cm]{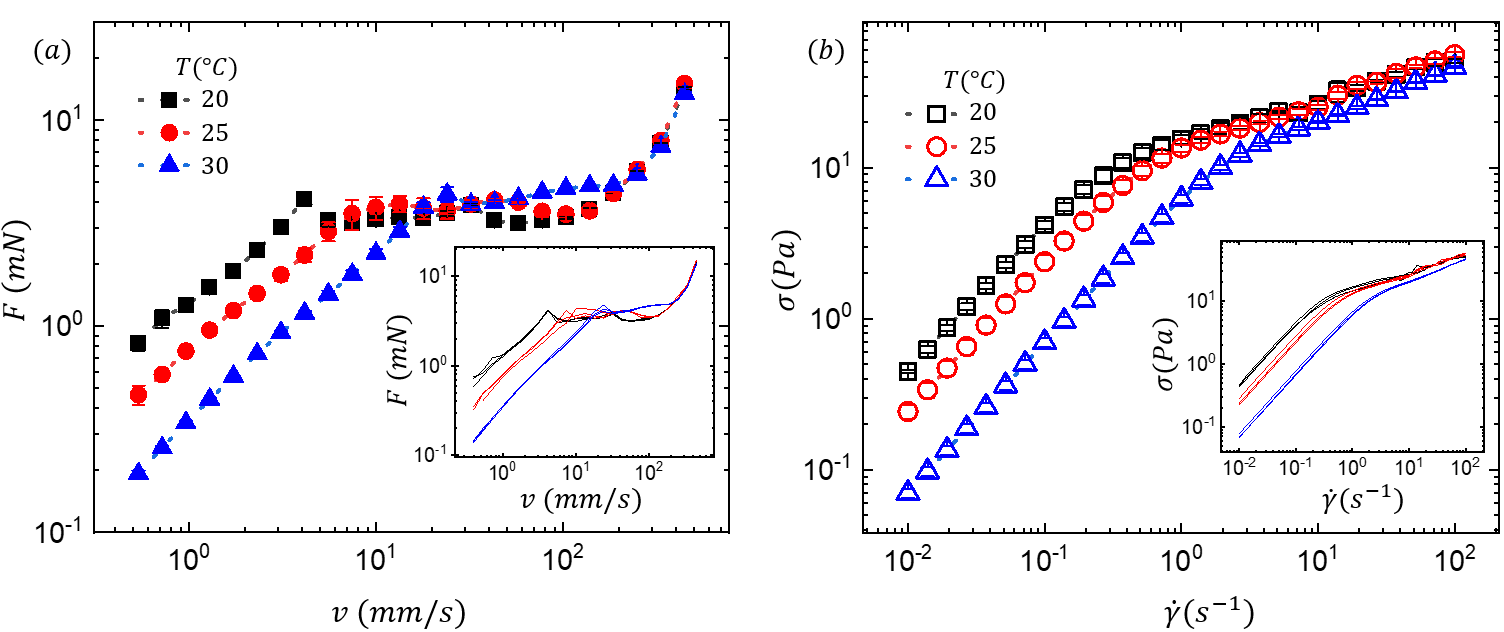}
\caption{\label{fig:wide} Comparison of $F-v$ curve in the case of pin geometry having 5 mm pin diameter (panel (a)) with the flow curve in the case of Taylor-Couette geometry (panel (b)) for 2wt\% CTAT + 100 mM NaCl system with variation of temperature. Clearly, a force plateau is present in the case of pin geometry, whereas the stress plateau is absent in the case of Taylor-Couette geometry. All the individual data sets are shown in the inset. The data for a particular sample at different temperatures are shown in different colors.  The data in panels a and b are highly reproducible, as indicated in the inset and also by the fact that the error bars are comparable to the size of the symbols. }
\end{figure}

\begin{figure}
\centering
\includegraphics[height=6.5cm]{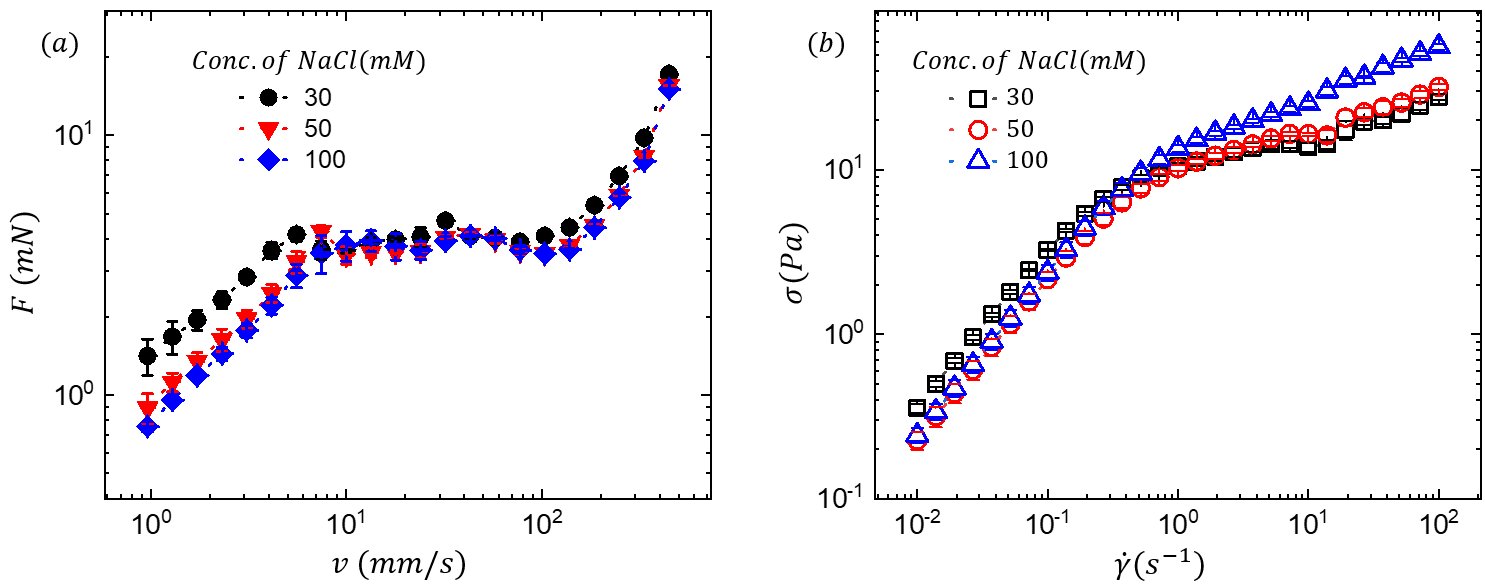}
\caption{\label{fig:wide} Comparison of $F-v$ curve in the case of pin geometry having 5 mm pin diameter (panel (a)) with the flow curve in the case of Taylor-Couette geometry (panel (b)) for 2wt\% CTAT with variation of NaCl concentrations at $25^{\circ}C$. Clearly, force plateau is present in the case of pin geometry whereas, the stress plateau is absent in the case of Taylor-Couette flow geometry.   }
\end{figure}

\begin{figure}
\centering
\includegraphics[height=7cm]{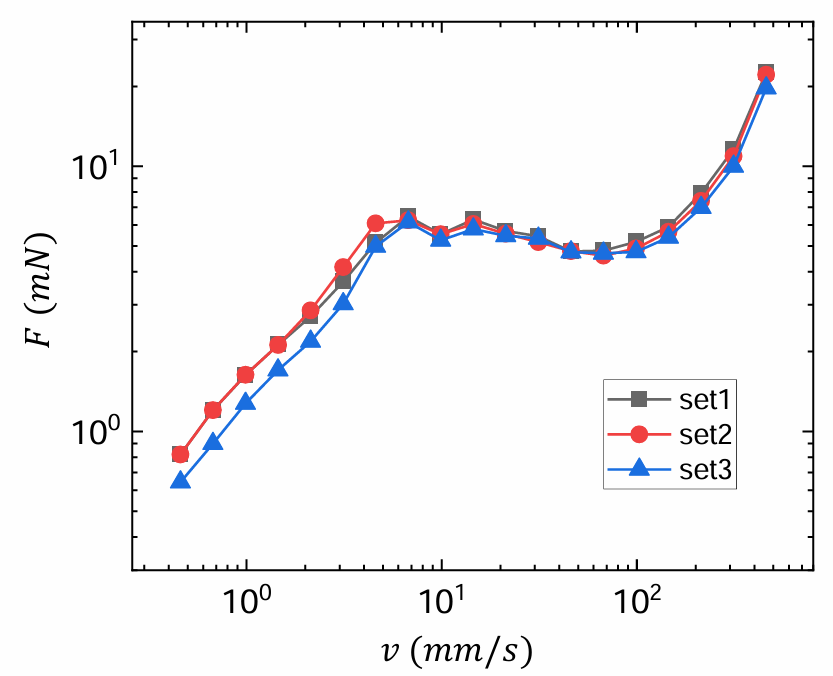}
\caption{\label{fig:wide} $F-v$ curve in the case of pin geometry for 2wt\% CTAT + 100 mM NaCl system at $25^{\circ}C$ for three different trials of the measurement. The data is reproducible over repeated measurements.   }
\end{figure}

\begin{figure}
\centering
\includegraphics[height=6.5cm]{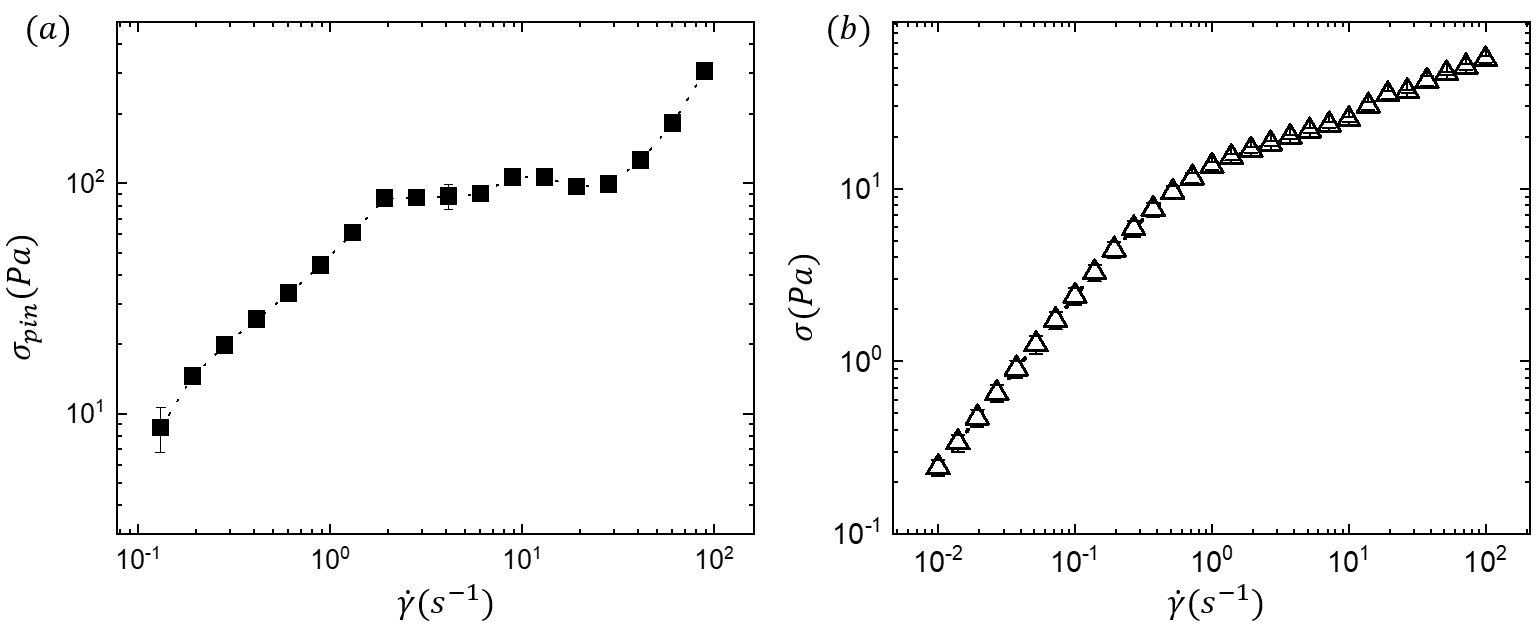}
\caption{\label{fig:wide} Stress due to pin ($\sigma_{pin}$)  and rate derived from force and velocity respectively in the case of pin motion (left panel) for 2wt\% CTAT + 100 mM NaCl at $25^{\circ}C$. Flow curve of the same sample obtained using a temperature-controlled Taylor-Couette geometry (right panel). Both flow curves are different by nature.  }
\end{figure}

\begin{figure}
\centering
\includegraphics[height=7.8cm]{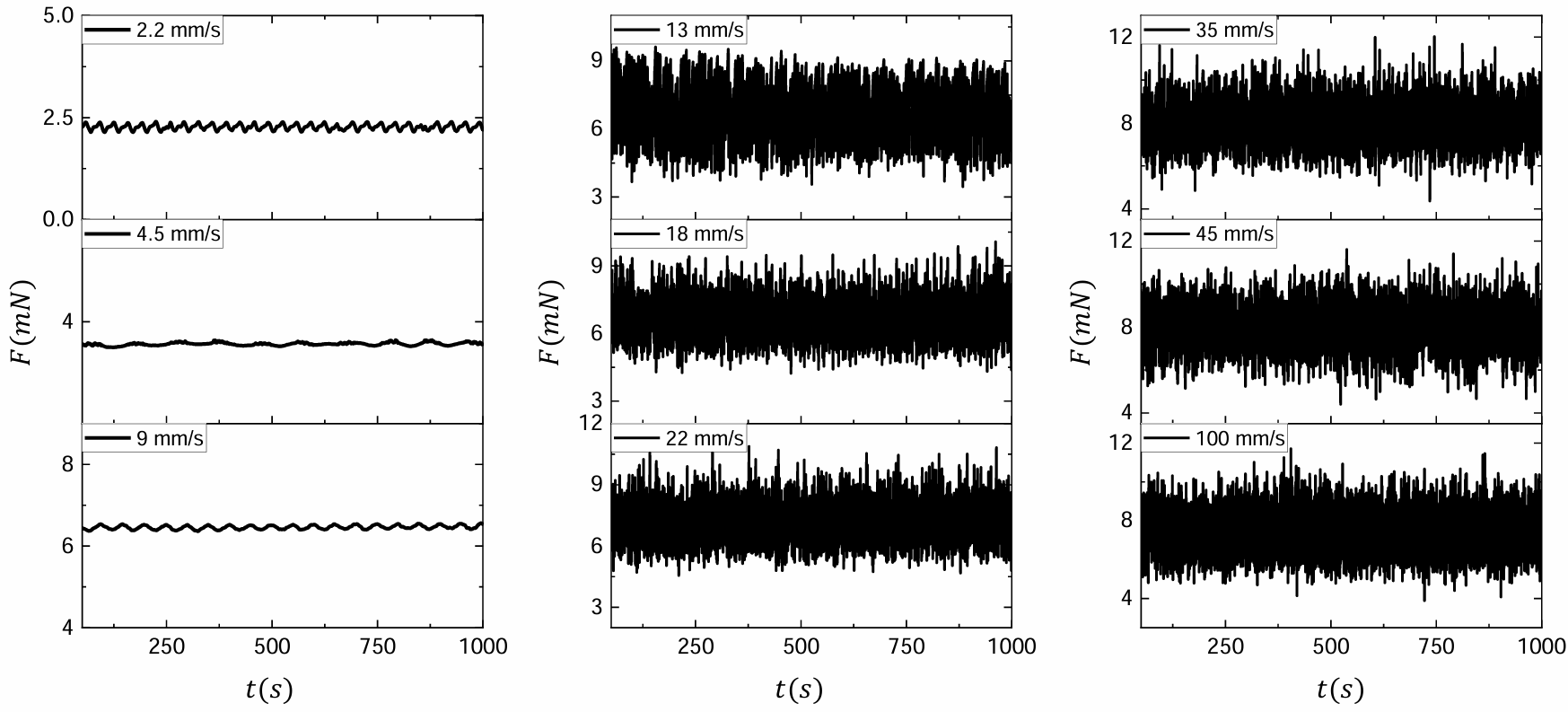}
\caption{\label{fig:wide} Temporal variation of force at different driving velocities (mentioned in the inset) for 2wt\% CTAT + 100 mM NaCl at $25^{\circ}C$ in the case of the  pin with 5 mm diameter. $F$ does not show temporal drift in any case and the amplitude of fluctuations remain consistent with time. }
\end{figure}

\begin{figure}
\centering
\includegraphics[height=7.8cm]{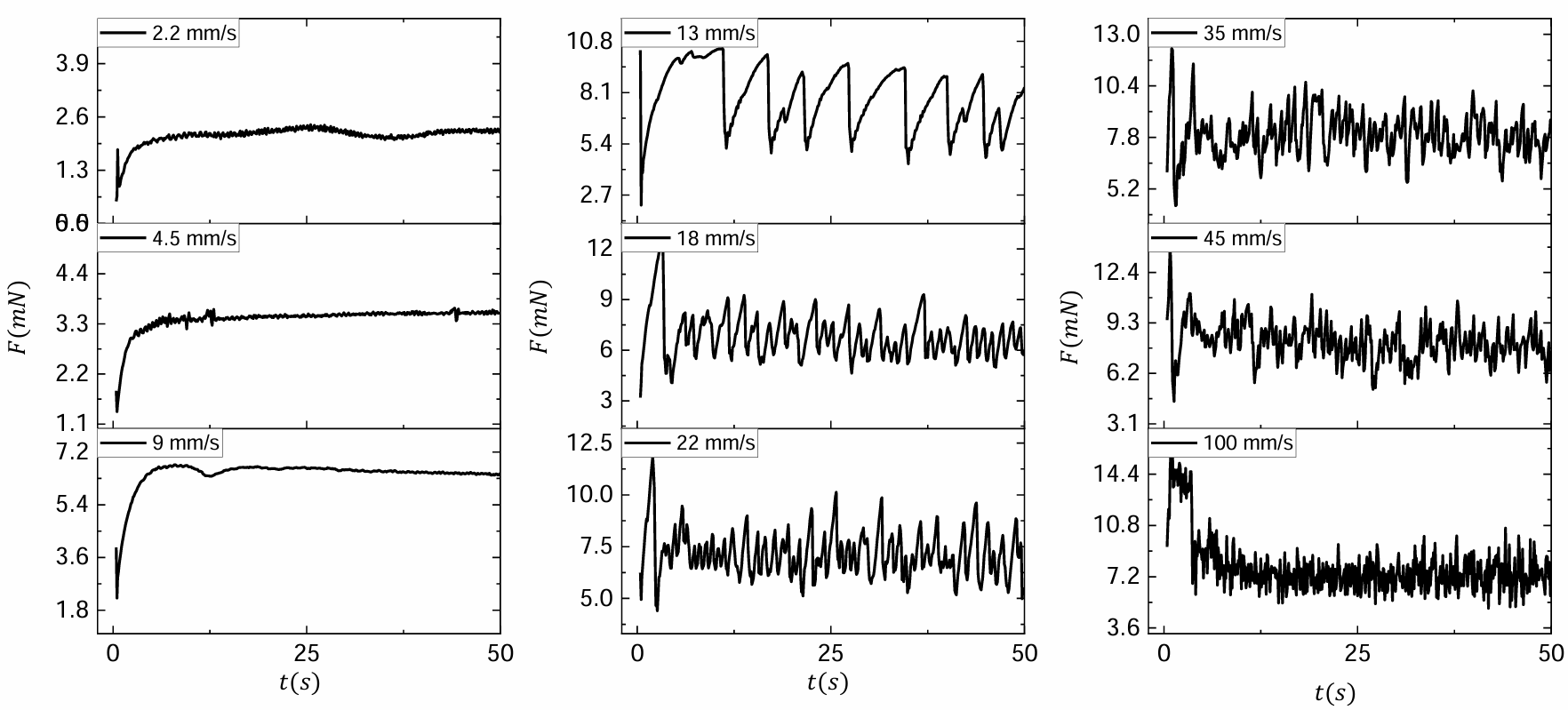}
\caption{\label{fig:wide} Variation of force at early times for different driving velocities (mentioned in the inset) for 2wt\% CTAT + 100 mM NaCl at $25^{\circ}C$ in the case of the  pin with 5 mm diameter. $F$ reaches steady state very quickly after the pin starts to move with a set velocity. }
\end{figure}

\begin{figure}
\centering
\includegraphics[height=8.5cm]{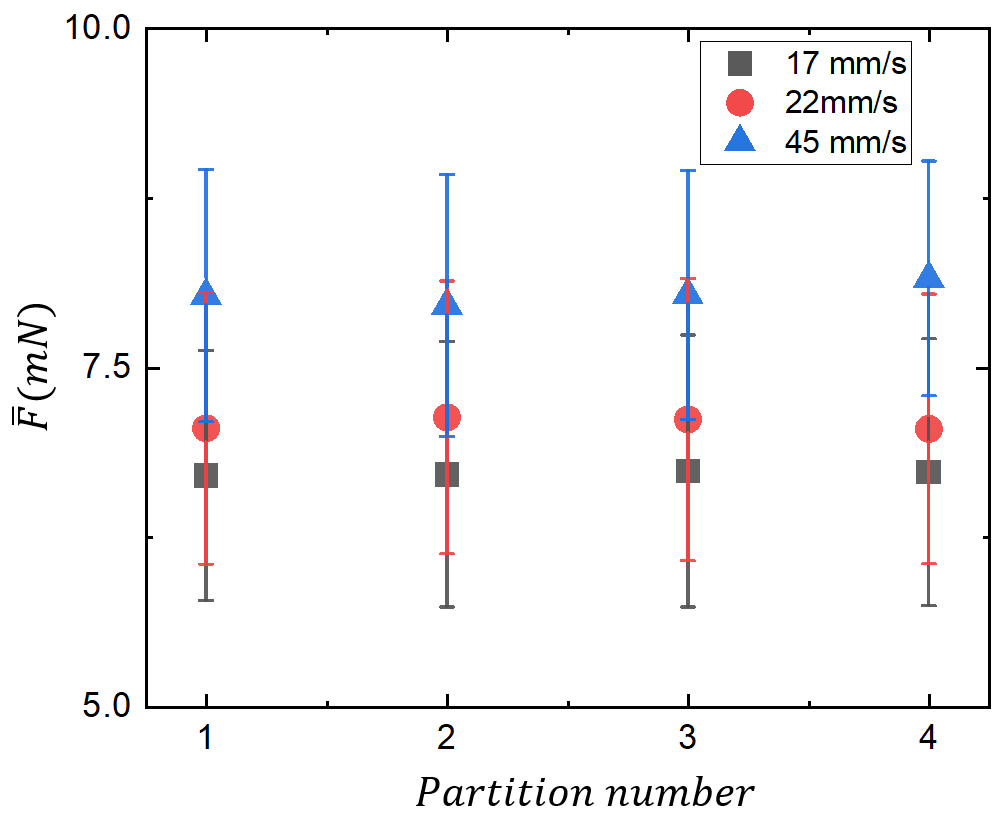}
\caption{\label{fig:wide} Mean value of force for different partition numbers obtained from the time series of force at different driving velocities (as indicated in the figure).
Each error bar represents a given partition's standard deviation of force fluctuations. The system does not show any temporal drift as indicated by almost constant mean force values and standard deviations across different partition numbers. }
\end{figure}

\begin{figure}
\centering
\includegraphics[height=9cm]{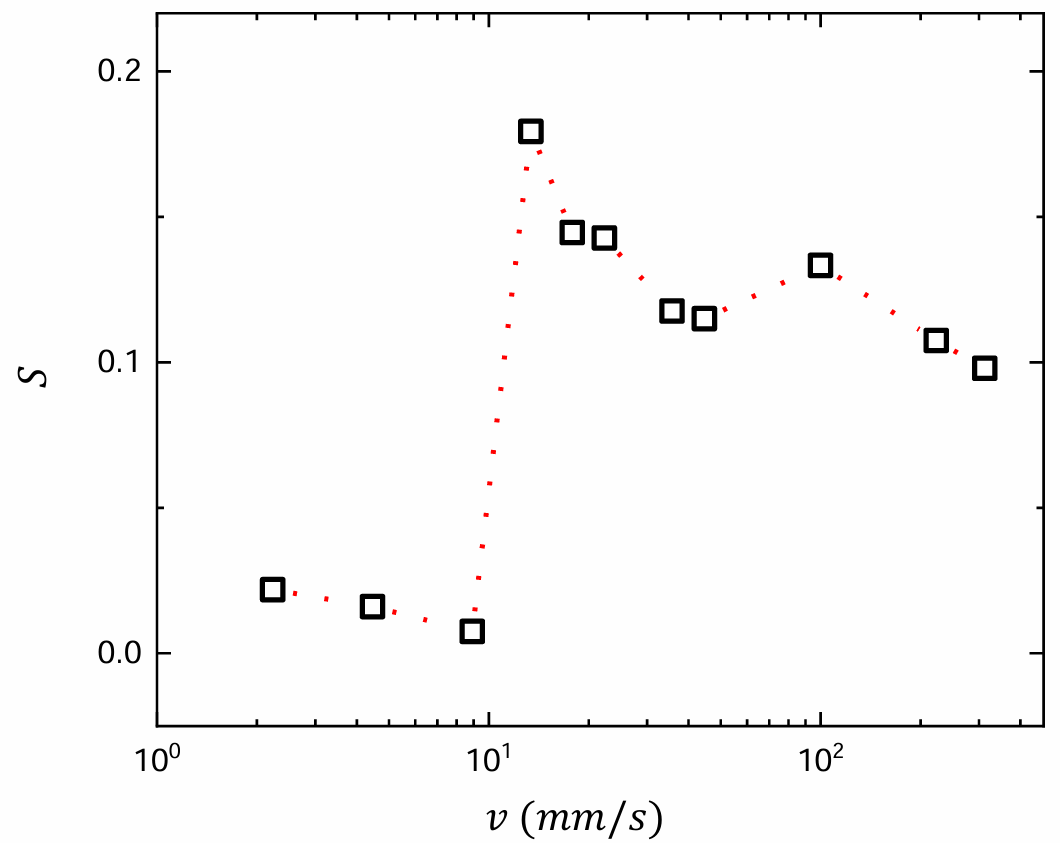}
\caption{\label{fig:wide} Ratio of standard deviation and mean of the temporal force fluctuation at different driving velocities (temporal fluctuations shown in Fig. S13). $S$ remains smaller for driving velocities that lies in the linear region of the $F(v)$ plot,  where $F$ does not show any significant temporal fluctuation. At the plateau onset, $S$ becomes maximum where the features of the fluctuation are of saw-tooth nature and decreases gradually in the deeper plateau region where the features of fluctuations become chaotic }
\end{figure}

\begin{figure}
\centering
\includegraphics[height=9.5cm]{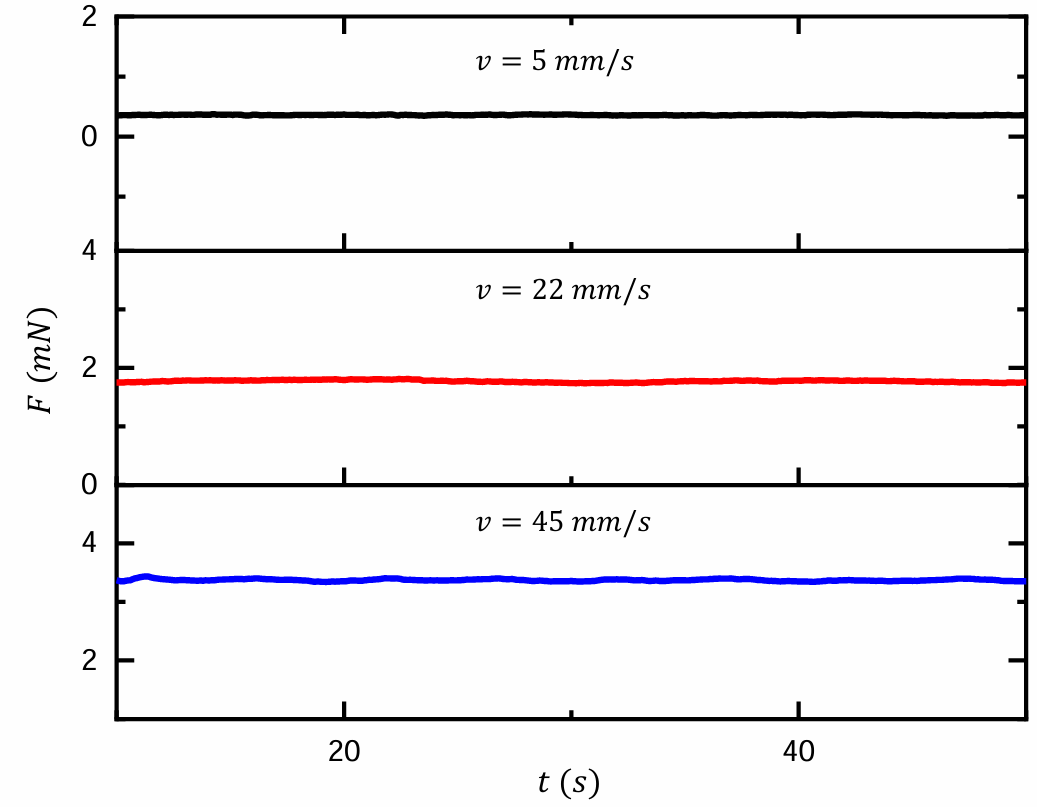}
\caption{\label{fig:wide} Temporal variation of $F$ at constant driving velocities of 5 mm/s, 22 mm/s, and 45 mm/s,  respectively (top to bottom) for Newtonian fluid glycerol showing no temporal fluctuations of force. }
\end{figure}

\begin{figure}
\centering
\includegraphics[height=5.2cm]{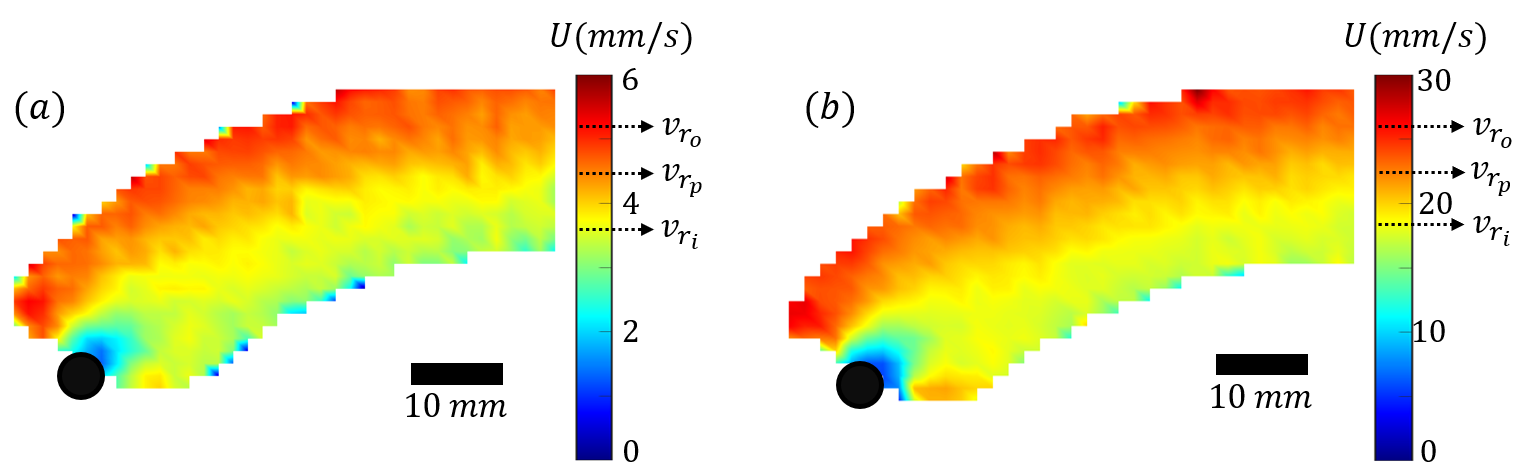}
\caption{\label{fig:wide} Spatial distribution of velocity magnitude obtained from PIV measurements for driving velocities of (a) 4.5 mm/s, and (b) 24 mm/s respectively for Newtonian fluid glycerol showing a very localized region of influence by the pin. Black circle represents the pin position and the colorbars show the variation of velocity magnitude. $v_{r_p}$, $v_{r_i}$, $v_{r_o}$ are marked as the driving velocity at the probe position, inner wall and outer wall of the sample cell respectively.}
\end{figure}

\begin{figure}
\centering
\includegraphics[height=12cm]{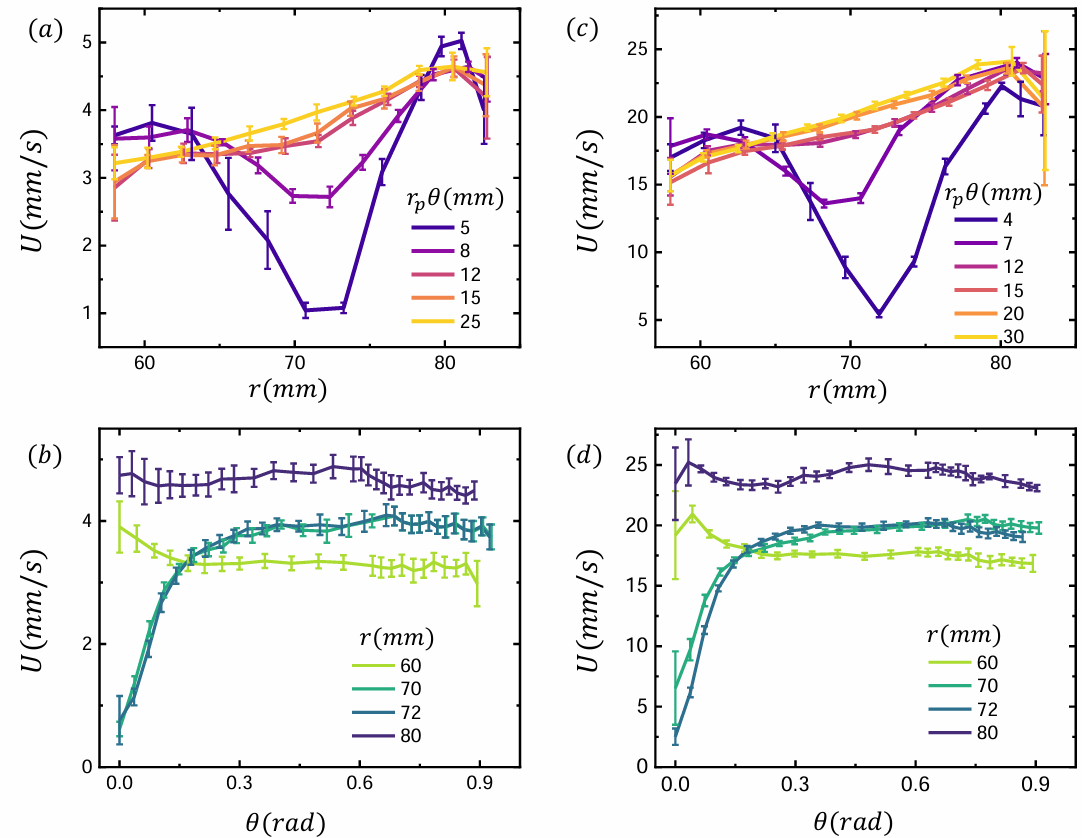}
\caption{\label{fig:wide} (a) and (b) are the radial and azimuthal variation in velocity respectively for a driving velocity of 4 mm/s for a Newtonian fluid glycerol obtained from PIV measurements. (c) and (d) are the radial and azimuthal variation in velocity for a driving velocity of 22 mm/s for the same liquid. The azimuthal ($r_p\theta$) and radial positions ($r$) are indicated by different colour lines.  In both the driving velocities, the region of influence remains same and local showing no dependence on the magnitude of driving velocities. However, this nature changes in the case of the micellar solution showing a much larger region of influence (Fig. 5 in the main text) due to the formation of a wake in contrast to glycerol. }
\end{figure}

\begin{figure}
\centering
\includegraphics[height=7.1cm]{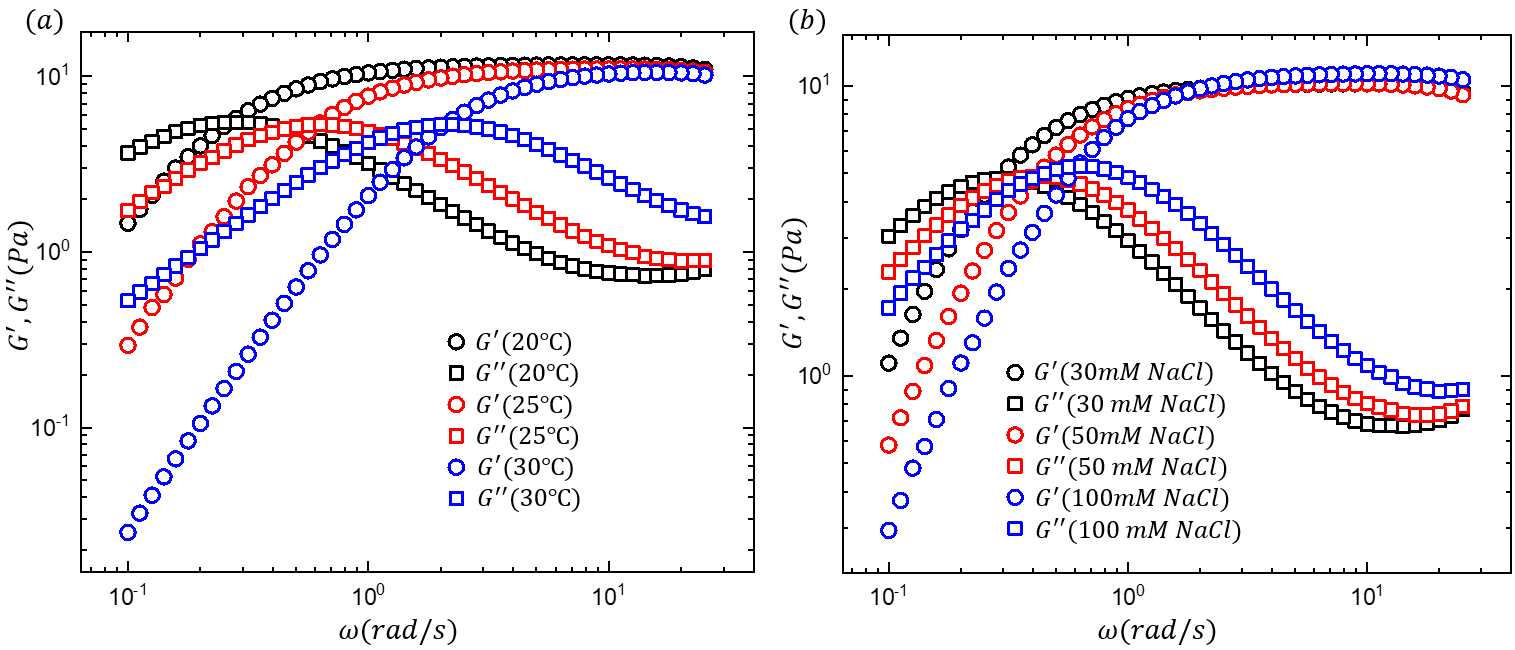}
\caption{\label{fig:wide} Storage($G'$) and loss($G''$) modulus as a function of angular frequency($\omega$) at a constant strain amplitude($\gamma$) of 5\% for 2wt\% CTAT + 100 mM NaCl system with variation of temperature(panel (a)). The shear rlaxation times obtained from the cross-over of $G'$ and $G''$ is found to be 4 s, 1.4 s, and 0.5 s for $20^{\circ}C$, $25^{\circ}C$ and $30^{\circ}C$ respectively. Similar measurements are done for 2wt\% CTAT system with variation of NaCl concentrations at $25^{\circ}C$.
 The shear rlaxation times obtained from the cross-over of $G'$ and $G''$ is found to be 3 s, 2 s, and 1.4 s for 30 mM, 50 mM, and 100 mM of NaCl respectively.}
\end{figure}

\begin{figure}
\centering
\includegraphics[height=6.5cm]{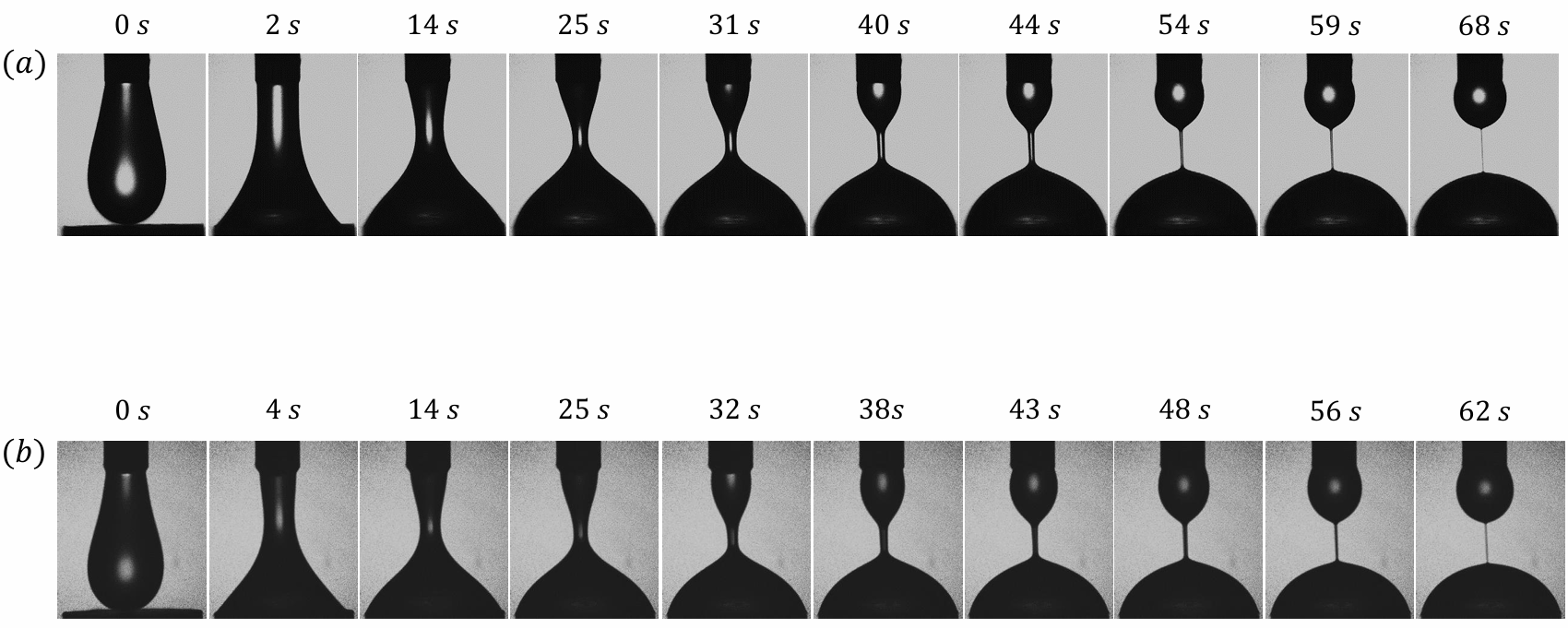}
\caption{\label{fig:wide} Experimental snapshots at different times for 2wt\% CTAT with 30 mM NaCl (panel (a)) and 50 mM NaCl (panel (b)) systems obtained from custom made DoS rheometry to probe the behaviour of the system under extensile deformation.}
\end{figure}

\begin{figure}
\centering
\includegraphics[height=9cm]{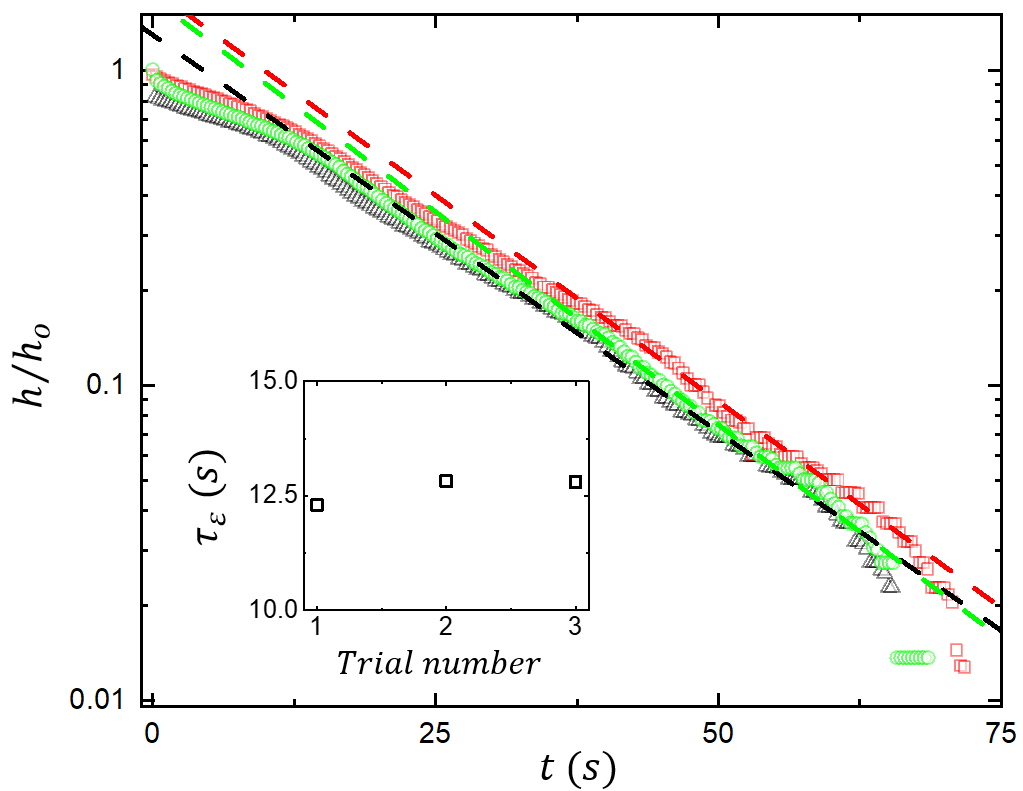}
\caption{\label{fig:wide} Temporal evolution of the minimum neck diameter of the liquid bridge normalized w.r.t. the needle diameter for 2wt\% CTAT + 30 mM NaCl system and the data shown for three repeated measurements marked as different colours. The dashed lines are the fit to the data using the elasto-capillary model in order to extract the extensional relaxation times. The variation of relaxation time obtained from DoS fittings over 3 trials is shown in the inset.}
\end{figure}

\end{document}